\documentclass[aps,11pt,onecolumn,preprintnumbers,amsmath,amssymb]{revtex4}

\usepackage[english]{babel}
\usepackage{graphicx}
\usepackage{epstopdf}
\usepackage{color}
\usepackage{ragged2e}
\usepackage{float}
\usepackage{xcolor}

\begin{document}

\newcommand{\unit}[1]{\:\mathrm{#1}}            
\newcommand{\To}{\mathrm{T_0}}
\newcommand{\Tp}{\mathrm{T_+}}
\newcommand{\Tm}{\mathrm{T_-}}
\newcommand{\EST}{E_{\mathrm{ST}}}
\newcommand{\Rp}{\mathrm{R_{+}}}
\newcommand{\Rm}{\mathrm{R_{-}}}
\newcommand{\Rpp}{\mathrm{R_{++}}}
\newcommand{\Rmm}{\mathrm{R_{--}}}
\newcommand{\ddensity}[2]{\rho_{#1\,#2,#1\,#2}} 
\newcommand{\ket}[1]{\left| #1 \right>} 
\newcommand{\bra}[1]{\left< #1 \right|} 

\bibliographystyle{naturemag}

\title{Optical control of valley Zeeman effect through many-exciton interactions}
\author{Weijie Li$^{\dagger}$}
\author{Xin Lu$^{\dagger}$}
\author{Jiatian Wu$^{\dagger}$}
\author{Ajit Srivastava$^{*}$}\affiliation{Department of Physics, Emory University, Atlanta 30322, Georgia, USA}
\maketitle
\justify
$^{\dagger}$These authors contributed equally to this work.\\
$^{*}$Correspondence to: ajit.srivastava@emory.edu

{\bf Charge carriers in two-dimensional transition metal dichalcogenides (TMDs), such as WSe$_2$, have their spin and valley-pseudospin locked into an optically-addressable index that is proposed as a basis for future information processing. The manipulation of this spin-valley index requires tuning its energy, typically through external magnetic field ($B$), which is cumbersome. Thus, other efficient routes like all-optical control of spin-valley index are desirable. Here, we show that many-body interactions amongst interlayer excitons in WSe$_2$/MoSe$_2$ heterobilayer induce a steady-state valley Zeeman splitting corresponding to $B$ $\sim$6 Tesla. This anomalous splitting, present at incident powers as low as $\mu$Ws, increases with power and enhances, suppresses or even flips the sign of a $B$-induced splitting. Moreover, the $g$-factor of valley Zeeman splitting can be tuned by $\sim$30\% with incident power. In addition to valleytronics, our results are relevant for achieving optical non-reciprocity using two-dimensional materials.}

The underlying honeycomb-lattice of group VIB semiconducting TMDs results in low-energy charge carriers possessing chirality which can be labeled by a pair of pseudospin indices, and identified with momentum-space $\pm K$-valleys. Consequently, circularly polarized light of a given helicity selectively couples to the valley with the corresponding chirality, as is observed in optical absorption or emission measurements~\cite{MakNNano2012, ZengNNano2012, CaoNComm2012}. Moreover, much as electron's spin, the valley-pseudospins are necessarily degenerate in the presence of time-reversal symmetry (TRS) and carry equal and opposite magnetic moments~\cite{XiaoPRL2007, XiaoPRL2012}. An out-of-plane $B$, couples to the valley magnetic moment and lifts the degeneracy of the $\pm K$ valleys. This magnetic control of valley-pseudospin, through valley Zeeman effect, has been well-established in monolayer~\cite{SrivastavaNPhys2015, AivazianNPhys2015, LiPRL2014, MacNeillPRL2015,Wang2DMaterials2015,StierNatComm2016,LyonsNatComm2019} and heterobilayer (hBL) TMDs~\cite{NaglerNComm2017,WangNL2019,CiarrocchiNP2019}. Even in the absence of an external $B$, an effective $B$ acting on the valleys can arise, for example, from a valley-contrasting optical Stark effect wherein a strong circular excitation, with photon energy typically below the absorption threshold, effectively breaks TRS in the TMD sample and causes a valley-splitting~\cite{KimScience2014, SieNMat2015}. 

In two-dimensional semiconducting TMDs, strong Coulomb attraction between an optically generated electron-hole pair leads to the formation of a tightly bound exciton~\cite{HePRL2014,ChernikovPRL2014}. The above-mentioned approaches for valley control are based on single-exciton effects which work by modifying the energies of the $\pm K$-valley excitons by the action of an external or effective $B$. In contrast, a $B$ can also arise through many-particle interactions when there is an imbalance in the densities of the two spin species. In the mean field picture of Stoner model of magnetism, an exchange field, arising from exchange interactions, can be thought of as an effective $B$ that each particle experiences due to the presence of all the other particles and is proportional to the spin imbalance density~\cite{BlundellMICM2003}. Whereas a valley-splitting by magnetic proximity effect has been observed~\cite{SanchezNL2016,ZhaoNNano2017,SeylerNL2018}, it is natural to ask whether an optically tuneable valley control based on many-exciton interactions can be realized in TMDs to expand the toolkit of valleytronics. Furthermore, as a valley imbalance can be achieved by optical means, this approach can be implemented in a dynamic and efficient manner.

Here, we experimentally investigate interlayer excitons (IXs) in heterobilayer (hBL) of WSe$_2$/MoSe$_2$ to address this question. The type-II band alignment of this hBL results in IXs with electrons (holes) confined in Mo (W) layer, which have a permanent electric dipole with a fixed orientation in the out-of-plane direction~\cite{CiarrocchiNP2019,LiNMat2020,KremserNPJ2D2020}. The exciton-exciton interaction amongst IXs can be approximated by two terms -- a valley-independent dipolar repulsion term, $U_\mathrm{dd}$, and a valley-dependent exchange interaction term, $U_\mathrm{ex}$, which raises (lowers) the energy of a parallel (an antiparallel) alignment of spin-valley indices~\cite{RiveraScience2016,LiNMat2020,KremserNPJ2D2020}. In a simple picture, we can understand the higher energy of the ferromagnetic alignment even in the presence of repulsive interactions due to the bosonic nature of excitons, unlike electrons.

\begin{figure}
\includegraphics[scale=0.39]{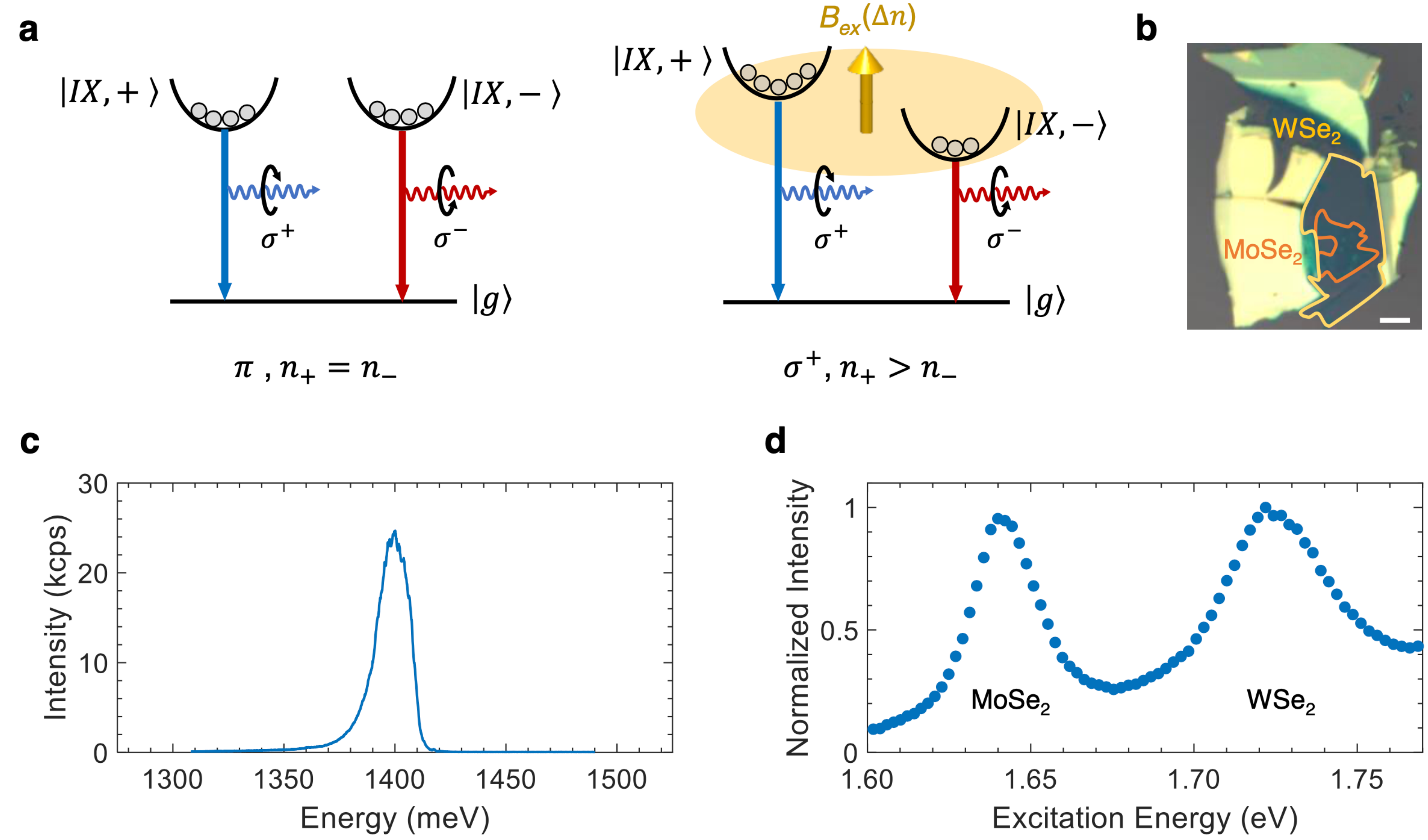}
\caption{{\bf Many-exciton exchange interactions amongst interlayer excitons of MoSe$_2$/WSe$_2$ heterostructure.} {\bf a,} Schematic of interlayer exciton (IX) valley energies under linear ($\pi$) and circular ($\sigma$) excitation. The valley-IXs emit $\sigma^+$ ($\sigma^-$) light in the state $|IX,+\rangle$ ($|IX,-\rangle$), and $|g\rangle$ is the exciton ground state. The populations of excitons, $n_\pm$, in the $\pm$K-valleys under $\pi$ excitation (left panel) are the same, while $\sigma^+$ excitation (right panel) induces an imbalance, $\Delta n = n_+ - n_- > 0$. The imbalance under  $\sigma^+$ excitation makes the exchange interaction between $|IX,+\rangle$ excitons larger when compared to $\pi$ excitation and gives rise to an effective exchange field  B$_\mathrm{ex}$($\Delta n$), shown by the yellow shaded region and arrow. {\bf b,} Optical microscope image of MoSe$_2$/WSe$_2$ heterobilayer sample. The WSe$_2$ (MoSe$_2$) flakes are outlined in yellow (orange). The scale bar is 5 $\mu$m. {\bf c,} Photoluminescence (PL) spectrum of the IX at 4 K, showing a strong peak at $\sim$ 1400 meV under 1 $\mu$W excitation power. {\bf d,} Photoluminescence excitation intensity plot, showing two prominent resonances 1.64 eV and 1.72 eV, corresponding to the monolayer MoSe$_2$ and WSe$_2$ intralayer exciton states. The intensity is integrated over the PL peak in {\bf c}. The excitation energy is 1.72 eV in {\bf c} and the excitation power is 2 $\mu$W in {\bf d}.} 
\end{figure}

In a many-exciton scenario, $U_\mathrm{dd}$ results in a exciton density-dependent blue-shift~\cite{LaikhtmanPRB2009,JaureguiScience2019} while $U_\mathrm{ex}$ results in an exchange-induced mean field ($B_{\mathrm{ex}}$) which depends on the imbalance, $\Delta n = n_{+} - n_{-}$, in the exciton densities ($n_{\pm}$) at the two spins or valleys~\cite{FernandezPRB1996,VinaPRB1996,AmandPRB1997, CiutiPRB1998, RiveraScience2016}. Fig.~1a demonstrates the basic idea behind our scheme based on IXs. When populations of $\pm K$ excitons are the same, TRS is unbroken and the two valleys remain degenerate. For a finite $\Delta n$, any given exciton experiences $B_{\mathrm{ex}}$ whose direction depends on the sign of $\Delta n$ such that its energy is raised (lowered) if it belongs to the valley with majority (minority) of excitons. In particular, the optically recombining excitons also experience $B_{\mathrm{ex}}$ and the resulting valley splitting can be measured in a helicity-resolved emission spectra.
%

Fig.~1b shows an optical microscope image of our fabricated heterostructure with monolayer MoSe$_2$ on top of WSe$_2$ with a very small twist angle between the two layers. Its photoluminescence (PL) spectrum at 4 K exhibits a peak at $\sim$1.40 eV (Fig.~1c), which is in the typical IX energy range~\cite{NaglerNComm2017,WangNL2019}. The strong emission intensity with an integrated (peak) intensity exceeding 3400 kCounts/s (25 kCounts/s) at a low excitation power of 1 $\mu$W, demonstrates the high quality of our sample (see Methods). In order to further confirm the interlayer nature of the peak, we conduct photoluminescence excitation (PLE) spectroscopy. The PLE spectrum shows two prominent resonances at 1.64 eV and 1.72 eV, corresponding to monolayer MoSe$_2$ and WSe$_2$ intralayer exciton states, as is expected for IX.

To optically create $\Delta n$, one can exploit the valley-contrasting optical selection rules for circular absorption in TMDs~\cite{XiaoPRL2012,RiveraScience2016}. By exciting with circularly polarized laser, say at the WSe$_2$ exciton resonance, $\Delta n$ can be efficiently created first in the WSe$_2$ layer which should then get transferred to the long-lived ($\sim$ns) IX on a very short timescale ($<$ 50 fs)~\cite{HongNNano2014,ZhuNL2017}. The spin-valley locking and the quenching of contact-type electron-hole exchange interaction in IXs because of the spatial separation of electron and hole in different layers is expected to suppress any valley-mixing during the relaxation of intralayer exciton to IX. This, together with the long IX lifetime~\cite{RiveraScience2016,TranNature2019}, should lead to an efficient generation of $\Delta n$ in the steady-state even at low excitation powers.

To test our claim about $B_{\mathrm{ex}}$, we first need to confirm the generation of  $\Delta n$ between $|IX, +\rangle$ and $|IX,-\rangle$ exciton densities, where $|IX, \pm \rangle$ denote the IXs in the $\pm K$-valleys. The imbalance can be characterized by the valley polarization or the degree of circular polarization (DCP) of PL, which is defined as $(I_\mathrm{co}-I_\mathrm{cross})/(I_\mathrm{co}+I_\mathrm{cross})$, where $I_\mathrm{co}$ ($I_\mathrm{cross}$) is the intensity of the co-polarized (cross-polarized) emission peak under circularly polarized excitation. As the DCP strongly depends on the excitation energy, we conduct PLE spectroscopy to decide the optimal excitation energy for generating valley imbalance. Fig.~2a shows that only excitation close to the WSe$_2$ resonance (1.72 eV) can create large positive DCP, while MoSe$_2$ resonance (1.64 eV) produces negligible DCP. This large positive DCP for WSe$_2$ resonance indicates that IX is co-polarized with the excitation helicity i.e., co-polarized excitons have much higher density leading to a large imbalance between two valleys. Although in the monolayer case, WSe$_2$ exciton shows a large valley polarization~\cite{AivazianNPhys2015} while MoSe$_2$ exciton can only have a small valley polarization~\cite{WangAPL2015}, one expects that at resonant excitation the latter should also exhibit a finite DCP~\cite{TornatzkyPRL2018}. The fact that IX emission hardly shows any DCP when excited at MoSe$_2$ resonance possibly hints at substantial valley mixing during the relaxation from MoSe$_2$ exciton to IX (see Supplementary section 1).

 \begin{figure}
\includegraphics[scale=0.3]{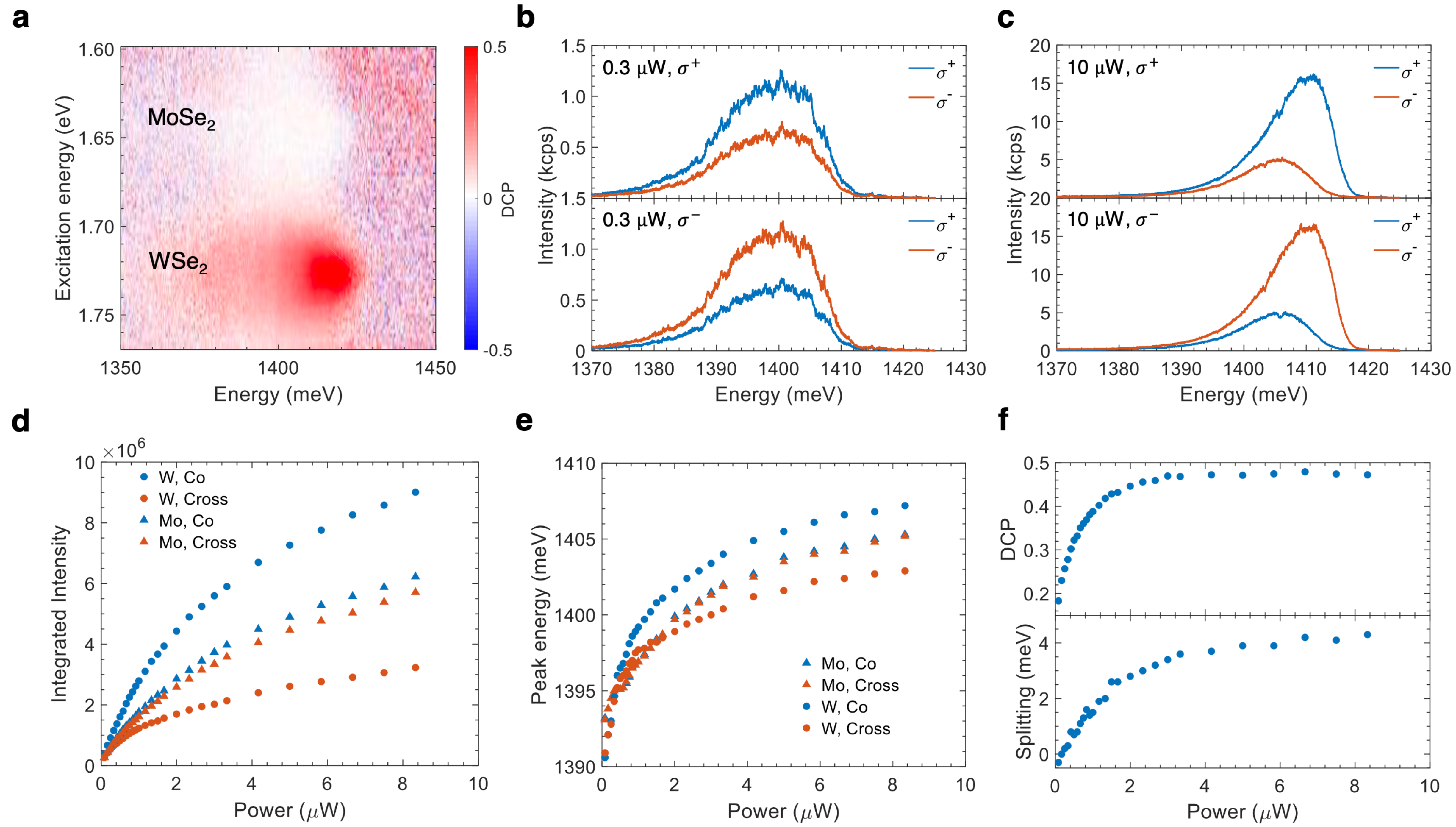}
\caption{{\bf Exchange field-induced splitting  in MoSe$_2$/WSe$_2$ heterostructure.} {\bf a,} The excitation energy dependence of the degree of circular polarization (DCP) defined as $(I_\mathrm{co}-I_\mathrm{cross})/(I_\mathrm{co}+I_\mathrm{cross})$, where $I_\mathrm{co}$ ($I_\mathrm{cross}$) is the intensity of the co-polarized (cross-polarized) interlayer exciton (IX). An excitation at WSe$_2$ (MoSe$_2$) resonance $\sim$1.72 eV (1.64 eV) creates a large (negligible) DCP, implying a large (negligible) imbalance between two valley-IX populations. {\bf b, c,} Polarization resolved photoluminescence (PL) spectra of interlayer excitons under low (high) excitation power of 0.3 $\mu$W (10 $\mu$W)  shown in {\bf b} ({\bf c}). The sample is excited with $\sigma^+$ ($\sigma^-$) light in the top (bottom) panel at 1.72 eV. The $\sigma^+$ ($\sigma^-$) component of the PL is shown in blue (red). At low power, ({\bf b}), no observable splitting between the $\sigma^+$ and $\sigma^-$ components is observed, while an obvious splitting is observed at high power ({\bf c}). The co-polarized peaks have higher intensity than the cross-polarized peaks.  {\bf d, e,} Power dependence of the integrated intensities and peak energies at the WSe$_2$ (MoSe$_2$) resonance denoted by circles (triangles). The co-polarized (cross-polarized) peak is shown in blue (red). At the WSe$_2$  resonance, the imbalance between intensities of co- and cross-polarized peaks and their peak energies increases with power, unlike for the MoSe$_2$ resonance. {\bf f,} Power dependence of DCP and splitting at the WSe$_2$ resonance. The splitting energy is $E_\mathrm{co}-E_\mathrm{cross}$, where $E_\mathrm{co}$ ($E_\mathrm{cross}$) is the energy of the co-polarized (cross-polarized) peak and follows the same trend as DCP, that is, increases with power and then saturates at high powers. The excitation power is 2 $\mu$W in the panel ({\bf a}). } 
\end{figure}

Next, we control $\Delta n$ by varying the intensity of circular excitation resonant with WSe$_2$ exciton and perform helicity-resolved PL spectroscopy at low power (0.3 $\mu$W, Fig.~2b) and high power (10 $\mu$W, Fig.~2c) with a laser spot-size of $~\sim$ 1 $\mu$m. With $\sigma^+$ ($\sigma^-$) excitation, the $\sigma^+$ ($\sigma^-$) emission is more intense, highlighting the co-polarized behavior mentioned earlier. At low circular power, the co-polarized emission has the same energy as the cross-polarized one within the spectral resolution. On the other hand, at high circular power where a large $\Delta n$ is expected, the co-polarized emission blue-shifts compared to the cross-polarized peak, regardless of the helicity of the excitation laser. On the other hand, the PL spectra under linearly polarized excitation does not result in any splitting and falls in between the two circular excitation spectra (see Supplementary section 1). In other words, circular excitation effectively breaks TRS and leads to an anomalous valley splitting at zero external $B$. The co-polarized emission with higher intensity has higher energy, consistent with the effect of $B_{\mathrm{ex}}$ shown in Fig.~1a. The zero-field splitting at a modest continuous-wave power of 10 $\mu$W is $\sim$4.5 meV, which based on the IX $g$-factor discussed below, is equivalent to a $B$ $\sim$6 Tesla. 

Our observations should be contrasted with a recent report of zero-$B$ valley-splitting observed in a similar TMD hBL wherein the lower intensity, cross-polarized peak shifts to a higher energy and the splitting only decreases with increasing circular power~\cite{JiangPRB2018}. Such a behavior is qualitatively different from our observations and inconsistent with exciton interactions induced valley-splitting but arises from an asymmetry in valley relaxation times of electrons and holes. On the other hand, our findings are similar to the helicity-induced Zeeman splitting of excitons in GaAs quantum wells which also originates from many-exciton interactions~\cite{FernandezPRB1996,VinaPRB1996,AmandPRB1997}. However, unlike our case, the splitting lasts several picoseconds under pulsed laser excitation and is absent in steady-state. We also remark that a laser intensity as low as 100 W/cm$^2$ is required to observe a valley-splitting of 1 meV in our scheme as opposed to $\sim$GW/cm$^2$ required to obtain a similar splitting using valley-contrasting optical Stark effect~\cite{KimScience2014, SieNMat2015}.

We perform a systematic power dependence under circular excitation at both WSe$_2$ and MoSe$_2$ resonances. As the power increases, the integrated intensities increases and saturates (Fig.~2d) and peak energies are blue-shifted (Fig.~2e). The saturation of the total intensity with power possibly arises from exciton-exciton annihilation (see Supplementary section 2). The blueshift results from both U$_\mathrm{dd}$ and U$_\mathrm{ex}$ between IXs. We note that the integrated intensity for MoSe$_2$ resonance are almost the same for the co- and cross-polarized emissions, and lie between the two WSe$_2$ resonance branches (Fig.~2d). If we assume that the integrated intensity is proportional to the exciton density, we can conclude that exciting at MoSe$_2$ resonance results in negligible valley imbalance even at higher powers. Indeed, this is consistent with the peak energies for MoSe$_2$ resonance being the same for the co- and cross-polarized peaks and falling in between the diverging peak energies for WSe$_2$ resonance (Fig.~2e). For WSe$_2$ resonance, we convert the difference in the integrated intensities (peak energies) into DCP (splitting) as shown in Fig.~2f (see Supplementary section 3 for detailed excitation energy dependence). The DCP increases from 20\% to 50\% and saturates beyond 3 $\mu$W, with a similar trend for the splitting which increases from 0 to $\sim$4.5 meV (see Supplementary section 4 for data on another IX). Using a theoretical model based on exciton-exciton interactions (see Supplementary section 5), the calculated splitting from peak shifts and DCP reproduce the experimental results fairly well, supporting the fact that the splitting arises from the imbalance between  $|IX,+\rangle$ and $|IX,-\rangle$. Assuming a binding energy, $E_\mathrm{b}$ = 200 meV, and a Bohr radius, $a_\mathrm{B}$ = 2 nm, we estimate that a splitting of $\sim$ 4 meV arises from a $\Delta n$ $\sim$ 3.3 $\times$ 10$^{11}$ cm$^{-2}$. This estimate agrees well for an incident power in the $\mu$W range with an absorption of $\sim$ 10\% at WSe$_2$ resonance and IX lifetime $\sim$ ns (see Supplementary section 5). We extract the the strength of exciton-exciton interaction from our experiments to be $\sim$0.8 $\mu$eV $\mu$m$^2$ which is about an order of magnitude larger than the previous studies on monolayer TMDs~\cite{TanPRX2020,BackPRL2018,ScuriPRL2018,BarachatiNNano2018} and is comparable to GaAs quantum well excitons~\cite{FernandezPRB1996,VinaPRB1996,AmandPRB1997}. Owing to their longer lifetime, a steady-state $\Delta n$ is efficiently created in TMD hBL as opposed to GaAs quantum wells.


\begin{figure}
\includegraphics[scale=0.3]{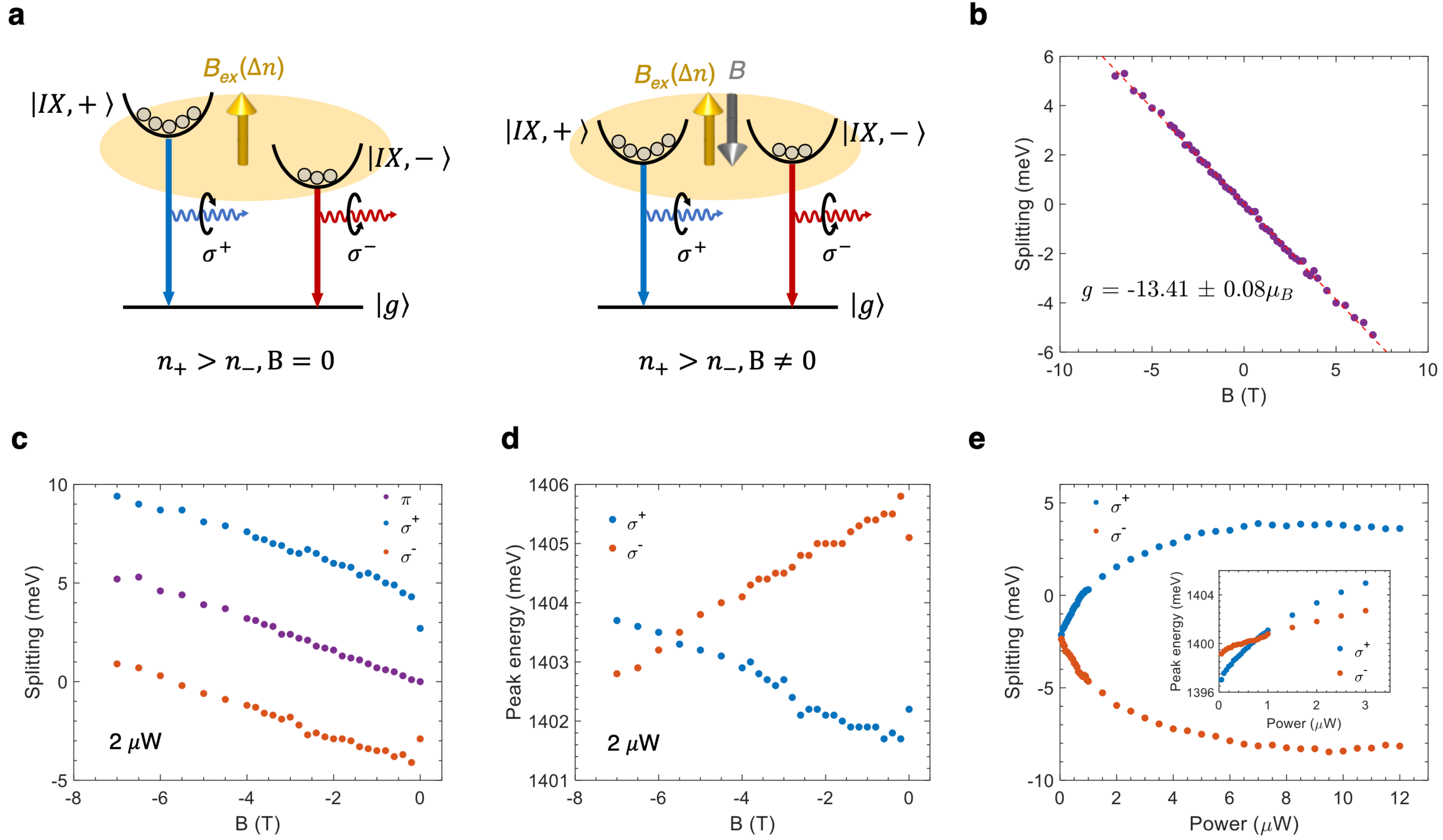}
\caption{{\bf Equivalence between the exchange field and the external magnetic field.} {\bf a,}  Schematic of valley-interlayer exciton (IX) energy levels at magnetic field ($B$) = 0 and $B$ $\neq$ 0 under $\sigma^+$ excitation. At $B$ = 0 (left panel), the circular excitation creates an exchange field, $B_\mathrm{ex}$, (yellow arrow) that lifts the degeneracy of the two valley-IXs. By applying an external $B$ (right panel, gray arrow), $B_\mathrm{ex}$ can be cancelled. {\bf b,}  Magnetic field dependence of the Zeeman splitting energy ($E_{\sigma^+}-E_{\sigma^-}$) under linear excitation. The $\it{g}$ factor of -13.41 is consistent with $60^\circ$ stacking angle. The pink dashed line is the linear fitting of the data. {\bf c,}  $B$ dependence of the splitting energy under circular excitation. At an excitation power of 2 $\mu$W, the splitting energy of the circular excitation ($E_{\sigma^+}-E_{\sigma^-}$) is equal to the $B_\mathrm{ex}$ induced splitting plus the Zeeman splitting under linear excitation. The violet, blue and red circles represent linear, $\sigma^+$, and $\sigma^-$ excitation. {\bf d,}  $B$ dependence of the peak energies under $\sigma^-$ excitation. The energy of the $\sigma^+$ (blue circles) and $\sigma^-$ components (red circles) are flipped by the external B field of -6 T. Thus, $\sigma^-$ excitation is equivalent to a positive $B_\mathrm{ex}$. {\bf e,} Power dependence of the splitting energies under circular excitation at $B$ = 3 T. At low (high) powers, the splitting is roughly equal to (larger than) the Zeeman splitting under the linear excitation. The inset shows the peak energy shift with the excitation power of $\sigma^+$ excitation, showing a flip $\sim$ 0.8 $\mu$W, which indicates that  $\sigma^+$  excitation is equivalent to a negative $B$. The excitation energy is 1.72 eV in all the panels.}
\end{figure}

To further investigate the analogy between $B_{\mathrm{ex}}$ and an external $B$, we perform magneto-PL spectroscopy. As shown in Fig.~3a, $\sigma^+$ pumping induces the imbalance between $|IX,+\rangle$ and $|IX,-\rangle$ at $B$ = 0, and thus the energy of $|IX,+\rangle$ is higher than that of the $|IX,-\rangle$. When an external $B$ is applied perpendicular to the sample ($B\neq$ 0), it shifts the energies of the two valleys in opposite directions by the valley Zeeman effect~\cite{NaglerNComm2017,WangNL2019,CiarrocchiNP2019}. The shift direction depends on the out-of-plane $B$ direction, and thus one expects that an external $B$ can cancel $B_\mathrm{ex}$ in one direction and enhance it in the other. When the $B_{\mathrm{ex}}$ is cancelled, the $\sigma^+$ and $\sigma^-$ components have the same energy (Fig.~3a, right panel). To test this picture, we first characterize the Land\'{e} $g$-factor of our hBL by measuring the valley Zeeman effect under linearly polarized excitation. As shown in Fig.~3b, we measure a $g$-factor of -13.41, suggesting that the sample is stacked with a twist angle $\sim$60$^\circ$~\cite{NaglerNComm2017}. 




Next, we measure the $B$-dependence of the splitting ($E_{+}-E_{-}$) which is the difference in the peak energies ($E_\pm$) of the $\sigma^\pm$-components of the PL, under $\sigma^{\pm}$ excitation.  The excitation power is chosen to be $\sim$ 2 $\mu$W so as to avoid any effects of power saturation. Figure 3c shows that for $\sigma^+$ ($\sigma^-$) excitation the magnitude of splitting increases (decreases) from 0 to -6 T. Thus, the $B_{\mathrm{ex}}$ generated by $\sigma^+$ ($\sigma^-$) excitation acts in concert (opposition) with the negative $B$. We note that the ``dip" in the splitting near 0 T is reminiscent of a similar behavior in DCP of long-lived excitons in TMDs under tiny $B$~\cite{SmolenskiPRX2016,JiangNComm2018} (see also Supplementary section 6 for dip-behavior in DCP data). As it is not the main focus of this study, in the following we choose to focus on $B$-dependence away from this dip. From Fig.~3d we find that the effect of anomalous splitting at zero field is completely cancelled by the external $B$ $\sim$-6 T for $\sigma^-$ excitation and the energies of $|IX,+\rangle$ and $|IX,-\rangle$ are flipped beyond -6 T (Fig.~3d). Remarkably, the ability to optically undo the effect of $B$ up to 6 T with continuous-wave power of $\sim$$\mu$W is attractive for spin-valley control, which has not been previously observed~\cite{JiangPRB2018,FernandezPRB1996,VinaPRB1996,AmandPRB1997}.   

Another evidence for the equivalence between $B_\mathrm{ex}$ and external $B$ is shown in Fig.~3e where we fix the external $B$ at +3 T and vary the circular excitation power. At very low powers, $B_\mathrm{ex}$ is negligible and the splitting of -2.3 meV is the same as the linear Zeeman splitting at +3 T in Fig.~3b. When the power increases, the $\sigma^+$ excitation cancels the external $B$ at $\sim$0.8 $\mu$W. The inset of Fig.~3e clearly shows that a flip in the energies of the two valleys at a positive $B$ is caused by $\sigma^+$ excitation. Thus, we can conclude that $\sigma^+$ ($\sigma^-$) excitation results in $B_\mathrm{ex}$ acting as negative (positive) external $B$ and that the $B_\mathrm{ex}$ and the external $B$ are analogous as far as the splitting is concerned. Thus, our scheme shows that an external $B$ which is slow and cumbersome to change, can be modulated with very low optical powers on a timescale of tens of nanoseconds.
  
%
%
%
%
%


From Fig.~3c, it appears that the splitting under $\sigma^\pm$ excitation is merely shifted from the linear Zeeman splitting by $\sim\pm$ 4 meV, such that the $g$-factor is independent of the helicity of excitation. We can then ask the question whether the $B_\mathrm{ex}$ and the external $B$ act together in a linear fashion even at higher powers, i.e., whether the total splitting, in the presence of circular excitation and $B$, is simply a sum of zero-field splitting and the valley Zeeman splitting for linearly polarized excitation. To answer this question, we first study the $B$ dependence at low circular power of 1 $\mu$W. As shown in the Fig.~4a, at -7 T, the expected spectra (dashed lines), assuming a linear behavior obtained by shifting the 0.1 T data, to avoid the dip-behavior, by the corresponding linearly-polarized Zeeman splitting, matches very well with measured spectra (solid lines). Therefore, we can conclude that there is negligible nonlinear behavior in the splitting at low powers. On the other hand, as shown in Fig.~4b, at higher circular power (7.6 $\mu$W), the expected spectra at -7 T and the measured spectra have a small but systematic difference. We find that $\sigma^-$ ($\sigma^+$) excitation which should reduce (enhance) the splitting, reduces (enhances) the splitting more than expected, implying a larger magnitude of $B_\mathrm{ex}$. When the external $B$ is flipped, a similar behavior is seen with the roles of $\sigma^\pm$ interchanged. 

\begin{figure}
\includegraphics[scale=0.42]{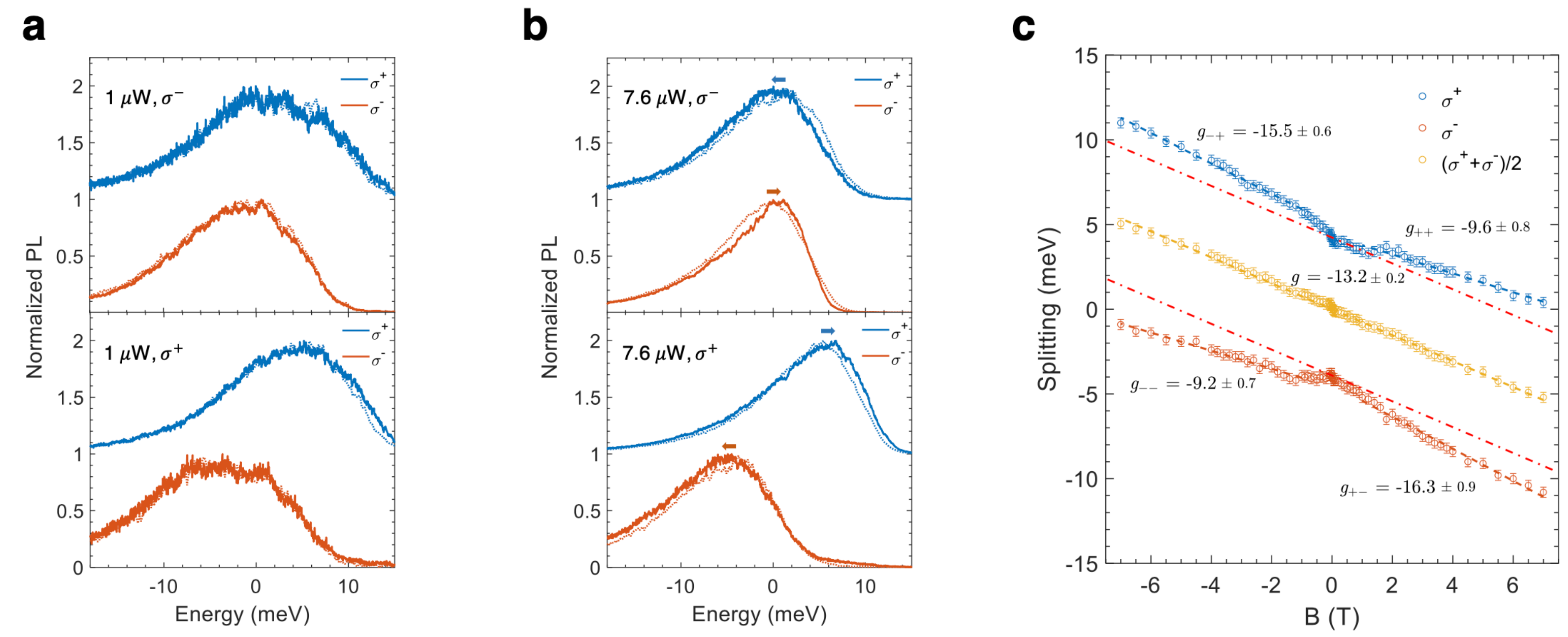}
\caption{{\bf Nonlinear behavior of Zeeman splitting under large circular-excitation power.} {\bf a, b,} Normalized photoluminescence (PL) spectra with an excitation power of 1 $\mu$W ({\bf a}) and 7.6 $\mu$W  ({\bf b}) at magnetic field ($B$) of -7 T. The top (bottom) panel shows spectra excited by $\sigma^+$ ($\sigma^-$) light. In order to show the splitting in a symmetric fashion, the zero of the energy is defined as the midpoint of fitted peak positions of $\sigma^+$ (blue) and $\sigma^-$ components (red). The blue and red dashed lines in {\bf a}, {\bf b} are the expected peaks at -7 T obtained by shifting the 0.1 T peaks by the corresponding Zeeman energy for linear excitation. At low circular power ({\bf a}), the experimental data matches with the expected curve, implying that the $g$-factor is independent of the excitation polarization. At high circular power ({\bf b}), $\sigma^+$ ($\sigma^-$) excitation enhances (cancels) the splitting more than expected as shown the left and right pointing arrows, implying a nonlinear behavior in power and $B$. The $\sigma^+$ emission is offset for clarity. {\bf c,} $\it{g}$-factor of different excitation polarization at an excitation power of 7.6 $\mu$W. $\it{g}$-factor of the $\sigma^+$ ($\sigma^-$) excitation under negative (positive) $B$ is - 15.5 (-16.3), which is larger than the linear case and consistent with {\bf b}. In contrast, $\sigma^-$ ($\sigma^+$) excitation under negative (positive) $B$ has a smaller $\it{g}$-factor. The blue, red and yellow dashed lines are linear fits for $\sigma^+$, $\sigma^-$ excitations, and the average of the two, excluding the dip near 0 T. The red dashed dot lines are parallel to the fitting curve of the averaged data which has a $g$-factor close to the linear case. The excitation energy is 1.72 eV in all the panels. } 
\end{figure}

To understand this nonlinear behavior further, we perform a systematic $B$-dependence and find that it leads to a helicity control of the bare $g$-factor (Fig.~4c). In particular, linear fitting of the $B$-dependence at 7.6 $\mu$W, away from the dip behavior near 0 T, gives a larger $\it{g}$-factor for $\sigma^+$ excitation under negative $B$ ($\it{g}_{-+}$ = -15.5) and $\sigma^-$ excitation under positive $B$ ($\it{g}_{+-}$ = -16.3). On the contrary, the $\it{g}$ factors of $\sigma^-$ excitation under negative $B$  ($\it{g}_{--}$ = -9.2) and $\sigma^+$ excitation under positive $B$ ($\it{g}_{++}$ = -9.6) are smaller. We note that even with the observed nonlinear behavior at higher circular power, the concept of a $g$-factor or linear-in-$B$ splitting remains valid up to at least $\pm$7 T. This behavior can be empirically captured by assuming that the magnitude of $B_\mathrm{ex}$ increases linearly with $|B|$ (see Supplementary section 7). Moreover, this increase in $|B_\mathrm{ex}|$ results from a corresponding increase in $|\Delta n|$. We make this conclusion by comparing the $B$-dependence of imbalance in emission intensities of the two helicities with that of the splitting. In other words, for a fixed, large incident circular power, increasing $|B|$ creates a larger $|\Delta n|$ (see Supplementary section 7). As the WSe$_2$ resonance shifts in $B$ only by $\sim$1 meV (see Supplementary section 7) while the PLE resonance is much broader (Fig.~1d), we rule out the possibility that this shift is responsible for the B-dependence of $\Delta n$. Moreover, as we excite on the WSe$_2$ resonance, even at 0 T, a large valley-imbalance between the initially created WSe$_2$ excitons is expected, which is unlikely to increase much more in $B$. Therefore, a possible explanation for this behavior could be that the valley-mixing processes during the relaxation of intralayer exciton to IX are suppressed with increasing $B$ and at higher circular powers. With a lower depolarization rate at a higher $B$, the imbalance increases, so does $B_\mathrm{ex}$, and thus induces a larger enhancement or cancellation of the splitting seen in Fig.~4b.

In conclusion, we have demonstrated an optical generation of exchange field in hBLs of semiconducting TMDs under steady-state condition. This effective magnetic field, arising from many-exciton interactions and essential for the control of the valley-pseudospin, can be dynamically tuned up to several Tesla. A combination of strong Coulomb interactions and characteristic optical properties of interlayer excitons in type-II hBL is responsible for the efficient generation of exchange field with very low continuous-wave incident powers ($\sim\mu$W) compared to previous schemes. In addition to valley-pseudospin control, our findings of TRS breaking in TMDs could lead to the engineering of gyrotropic medium in an energy-efficient manner for optical non-reciprocity~\cite{ShiNP2015}. On a fundamental side, our experiments should motivate the exploration of strongly interacting quantum phases of light and matter under driven-dissipative conditions~\cite{BarachatiNNano2018}. In particular, recent reports of exciton condensation and optical spectroscopy of strongly correlated electronic phases in moir\'{e} heterostructures of TMDs have garnered much attention~\cite{WangNature2019,TangNature2020,ReganNature2020,ShimazakiNature2020}. In light of this, our results on many-body physics of excitons, focusing on the valley-pseudospin, make a strong case for investigating quantum magnetism in this rich materials platform.

\justify

\textbf{Acknowledgments} We acknowledge many enlightening discussions with Ata\c{c} Imamo\u{g}lu.  A. S. acknowledges support from NSF through the EFRI program-grant \# EFMA-1741691 and NSF DMR award \# 1905809. 

\justify
\textbf{Author Contributions} ~$^\dagger$ W. L., X. L., J. W contributed equally to this work.  A. S., W. L., X. L. conceived the project. W. L., X. L., J. W. carried out the measurements.  J. W. prepared the samples. A. S. supervised the project. All authors were involved in analysis of the experimental data and contributed extensively to this work.

\justify
\textbf{Competing interests} The authors declare no competing interests.

\justify

\textbf{Methods}
\\
\textbf{Sample fabrication} 
We transfer the mechanically exfoliated samples by polydimethylsiloxane-based dry transfer method on 300~nm SiO$_2$/Si substrates. Monolayer MoSe$_2$ (HQ graphene) is stacked on top of monolayer WSe$_2$ (HQ graphene) with the aligned edges. After the stacking, the sample is annealed in 5$\%$ H$_2$/95$\%$ N$_2$ at 125$^{\circ}$C for 2~h. 
\justify
\textbf{Magneto-optical spectroscopy}
 We use two home-built, low temperature ($\sim$4~K) confocal microscope setups for magneto-photoluminescence measurements.
The sample is first loaded into a closed-cycle cryostat (AttoDry 800, base temperature $\sim$4K) for the sample characterization and subsequently into another cryostat (BlueFors cryogenics, base temperature $\sim$3.2K) for Faraday geometry measurements (from -8 T to +8 T).  The sample is positioned by a coarse and fine piezoelectric nanopositioners (Attocube systems).
The emission is collected using either a room or low temperature achromatic objective (NA = 0.42 for AttoDry 800 and NA = 0.63 for BlueFors cryogenics) and directed to a high-resolution (focal length: 500 mm for AttoDry 800 and 750 mm for BlueFors cryogenics) spectrometer (Princeton Instrument HR-500 for AttoDry 800 and Princeton Instruments SP-2750i for BlueFors cryogenics) where it is dispersed by a 1200 g/mm or 300 g/mm grating (both blazed at 750 nm). 
A charge coupled device (Princeton Instrument PIXIS-400 CCD for AttoDry 800 and PyLoN CCD for BlueFors cryogenics) is used as a detector.
The excitation laser is a mode-hop-free tuneable continuous-wave Ti:Sapphire laser (MSquared Lasers) with resolution of 0.1 pm, which is focused to a spot size of 1~$\mu$m on the sample. 
The polarization of incident laser is controlled by using a polarizer together with a liquid crystal variable retarder ($\lambda/4$ waveplate) for Bluefors cryogenics (AttoDry 800).
Polarization-resolved measurements are performed by using a $\lambda/4$ waveplate (achromatic, 690-1200 nm) placed before a Wollaston prism. 
The mechanism is that $\lambda/4$ waveplate transforms circular emission is converted into linearly polarized light  s- and p-components, and then the two components are displaced separately by the Wollaston prism. Another achromatic $\lambda/4$ waveplate is placed after the Wollaston prism to convert the linearly polarized light into a circular signal, in order to avoid the polarization sensitivity of the grating. 

\justify
\textbf{Data availability}
\\
The data that support the plots within this paper and other findings of this study are available from the corresponding author upon reasonable request.

\newpage
\textbf{\underline{Supplementary Information Contents:}}\\
Note 1. Discussion on negligible valley polarization on the MoSe$_2$ resonance (section 1).\\
Figure S1. Interlayer exciton configurations for WSe$_2$ and MoSe$_2$ resonances in AB-stacking heterobilayers\\
Figure S2. Dependence of emission energy and intensity on incident polarization (section 1).\\
Figure S3. Power dependence of the peak energy, degree of circular polarization and splitting at linear excitation (section 1)\\
Note 2. Estimation of the exciton-exciton annihilation rate from the power dependence of integrated intensity (section 2).\\
Figure S4. Calculated power dependence of the integrated intensity (section 2).\\
Figure S5. Excitation dependence of the peak energy and splitting at circular incidence (section 3).\\
Figure S6. Power dependence of  the degree of polarization and splitting for another interlayer exciton (section 4).\\
Note 3. Calculation of the power-dependent energy splitting (section 5).\\
Figure S7. Calculated power dependence of the energy splitting (section 5).\\
Figure S8. Magnetic field dependence of the degree of circular polarization (section 6). \\
Figure S9. Power dependence of the degree of circular polarization at zero and finite magnetic fields (B) (section 6).\\
Figure S10. Polarization resolved reflectance spectra at linear excitation at B = 7 T (section 7).\\
Note 4. Magnetic field dependence of exchange field ($B_{\mathrm{ex}}$) (section 7).\\
Note 5. Calculation of the magnetic field dependence of the energy splitting (section 7).\\
Figure S11. Relationship between the imbalance in intensities of the circular components to the splitting (section 7).\\


\clearpage
\justify

\textbf{Note 1: Discussion on negligible valley polarization on MoSe$_2$ resonance.}
\\
The negligible valley polarization of interlayer excitons at MoSe$_2$ resonance may be understood based on the their formation process. In the main text, Fig.~1d shows that the interlayer excitons have two resonances corresponding to WSe$_2$ and MoSe$_2$ resonances; however, Fig. 2a shows the robust conservation of valley polarization at WSe$_2$ resonance but negligible valley polarization at MoSe$_2$ resonance, which is also reported in former report\cite{BrotonsNMat2020}. The difference between two resonances are consistent with the fact that monolayer MoSe$_2$ shows less valley polarization than monolayer WSe$_2$\cite{WangAPL2015}. In other words, the intralayer valley depolarization behavior before the fast transfer of carriers between layers makes MoSe$_2$ resonance more depolarized compared to WSe$_2$ resonance. Nevertheless, it still can not explain the complete lost of valley polarization on MoSe$_2$ resonance, especially resonantly excited monolayer MoSe$_2$ showing valley polarization\cite{TornatzkyPRL2018}.

Another possible explanation is that the carriers are depolarized during the carrier transfer process at MoSe$_2$ resonance. According to the $\it{g}$ factor of -13.41, the twist angle of the heterobilayer is close to 60$^\circ$, i.e., AB stacking. The valley configuration is that K (-K) valley of WSe$_2$ is aligned with -K (K) valley of MoSe$_2$ as shown in Fig.~S1. When we pump the sample with $\sigma^+$ light at WSe$_2$ resonance (Fig.~S1a), electrons are excited in K valley of WSe$_2$ and then transferred to -K valley of MoSe$_2$ via spin-conserving process, forming singlet interlayer excitons. In contrast, $\sigma^+$ excitation at MoSe$_2$ resonance requires hole spin flip process to form the interlayer excitons, no matter for the singlet or triplet states (Fig.~S1b). The process that hole transfers to K valley of  WSe$_2$ is not considered because of the large momentum mismatch. The spin flip process at MoSe$_2$ resonance may lose the information of valley polarization and thus the interlayer excitons are completely depolarized. Since the optical selection rules are site dependent in the heterobilayers\cite{Yu2DMat2018}, from the fact that only one strong peak is observed and is co-polarized, one possibility is that the interlayer excitons are at $H_h^x$ site, where the observed excitons are in the singlet state with co-polarized feature, while the triplet states are coupled to z direction and unobserved. For the singlet states, WSe$_2$ resonance can create interlayer excitons without flipping spins, whereas the MoSe$_2$ resonance needs to either flip both the electron and hole spins or create the same configuration as Fig.~S1a through intervalley scattering.

\begin{figure}[H]
\centering
\includegraphics[scale=0.9]{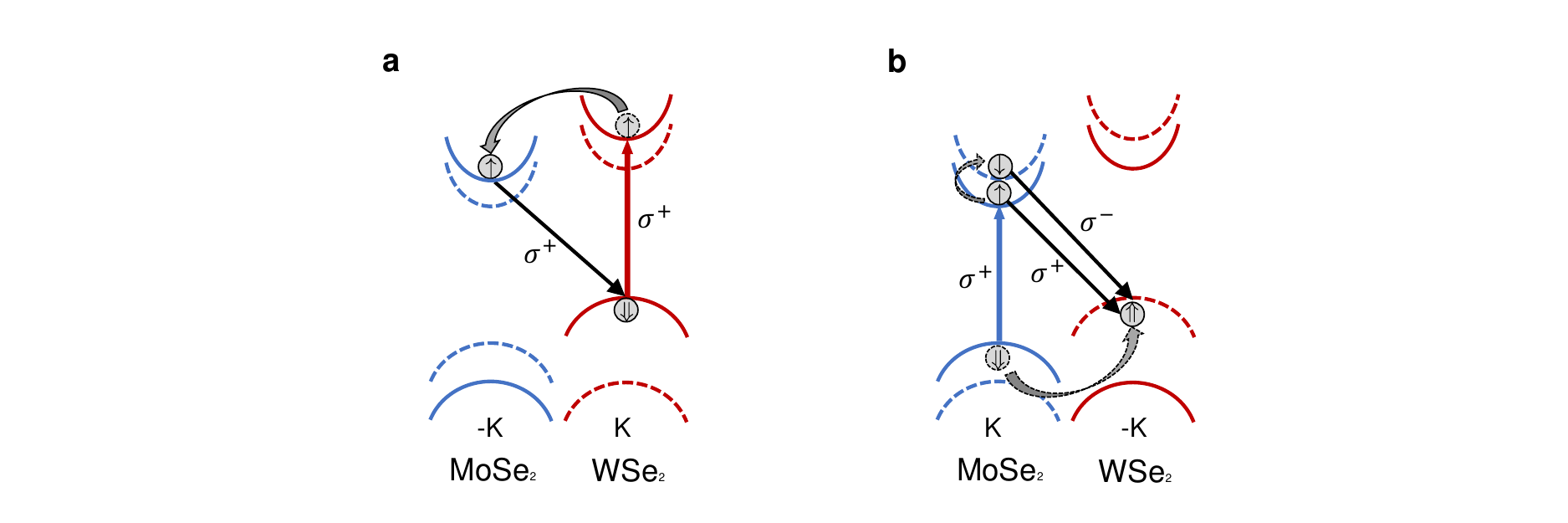}\\
\justify
{\bf Figure S1:  Interlayer exciton configurations for WSe$_2$ and MoSe$_2$ resonances in AB-stacking heterobilayers.}  {\bf a,} $\sigma^+$ excitation for WSe$_2$ resonance. For the heterobilayer with a twist angle of 60$^\circ$, K valley in the WSe$_2$ layer is aligned with -K valley in the MoSe$_2$ layer. When WSe$_2$ is excited by $\sigma^+$ light (red arrow), the spin-conserving tunneling of electrons (solid grey arrow) from WSe$_2$ to MoSe$_2$ forms $\sigma^+$-polarized interlayer excitons (black arrow). The solid (dashed) lines in the band structures are spin up (down) states. {\bf b,} $\sigma^+$ excitation for MoSe$_2$ resonance. Due to the time reversal symmetry, K valley in the MoSe$_2$ layer is aligned with -K valley in the WSe$_2$ layer. When $\sigma^+$ light excites at MoSe$_2$ resonance (blue arrow), the interlayer excitons must be formed via spin-flipped process (dashed grey arrow) or change the valley index (not shown), which may lose the information of polarization.
\end{figure}

\begin{figure}[H]
\centering
\includegraphics[scale=0.9]{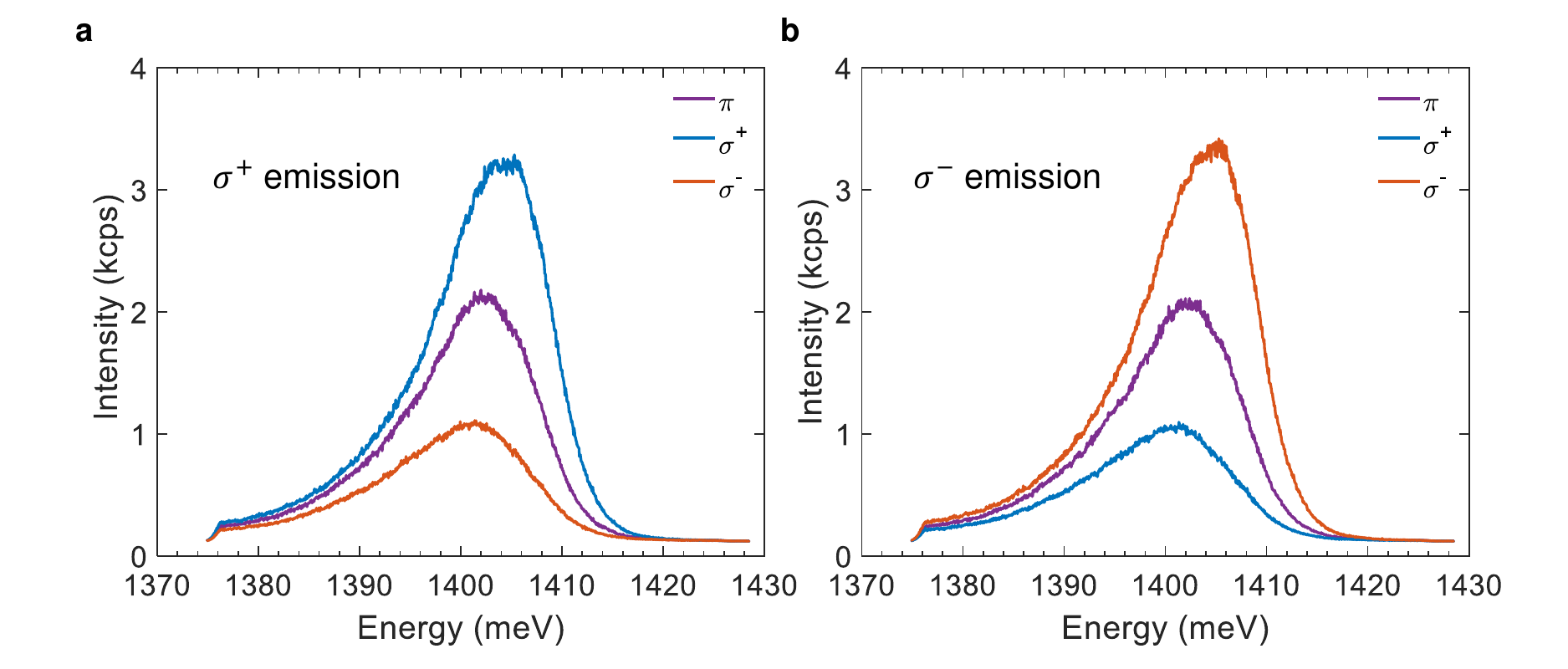}\\
\justify
{\bf Figure S2: Dependence of emission energy and intensity on incident polarization.} {\bf a,} The dependence of $\sigma^+$-polarized interlayer excitons. The $\sigma^+$-polarized interlayer excitons have the highest energy and intensity at $\sigma^+$-polarized excitation (co-polarized) and the lowest energy and intensity at $\sigma^-$-polarized excitation (cross-polarized). The linear excitation gives $\sigma^+$-polarized interlayer excitons the energy and intensity between the two cases, which is consistent with Fig.~1a in the main text. {\bf b,} The dependence of $\sigma^-$-polarized interlayer excitons. Due to the time reversal symmetry, $\sigma^-$-polarized interlayer excitons at linear excitation still have the energy and intensity between the co-polarized and cross-polarized excitation. The excitation energy is 1.72 eV with a power of 5 $\mu$W.
\end{figure}

\begin{figure}[H]
\centering
\includegraphics[scale=0.9]{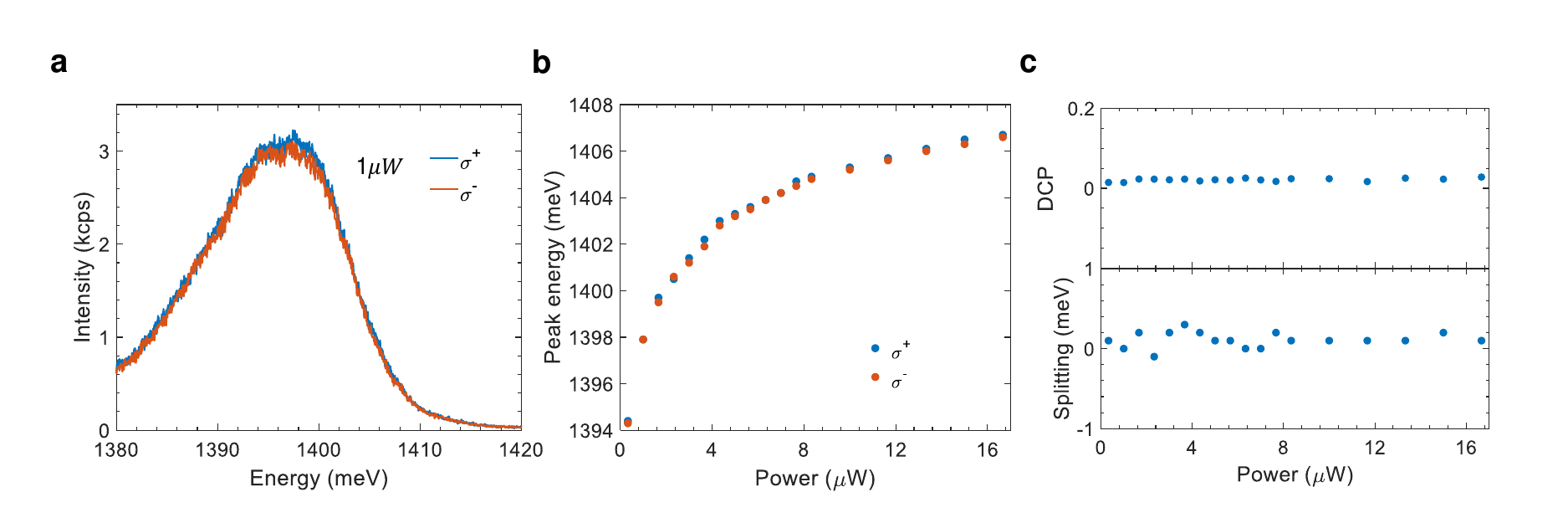}\\
\justify
{\bf Figure S3: Power dependence of the peak energy, degree of circular polarization and splitting at linear excitation} {\bf a,} Polarization resolved PL spectra at the linear polarized excitation. Linear polarized excitation induces no imbalance and no energy splitting between $\sigma^+$ and $\sigma^-$-polarized interlayer excitons.  {\bf b, c,} Power dependence of the peak energy shift (b), DCP (defined as $(I_+-I_-)/I_++I_-)$) in the upper panel of (c) and splitting in the bottom panel of (c). Peak energies of 
two interlayer exciton species blueshift with the increased power, but the energy splitting and imbalance remain to be zero over the power range. The excitation energy is 1.72 eV in all panels.
\end{figure}

\textbf{Note 2: Estimation of the exciton-exciton annihilation rate from the power dependence of integrated intensity.}
\\
Saturation of the power dependent integrated intensities is an important factor to consider, as it may account for the saturated splitting at high power. Two kinds of possible mechanisms are discussed here, exciton-exciton annihilation (EEA) and absorption saturation. The EEA is mainly non-radiative Auger recombination, which has been observed in various transition metal dichalcogenide monolayers \cite{MouriPRB2014, KumarPRB2014, SunNL2014, YuPRB2016, HoshiPRB2017, KuligPRL2018}. This process has the probability of $kn^2$, where $k$ is the Auger coefficient and $n$ is the exciton density. Therefore, the rate equation for EEA under continuous wave excitation is given by,

$$\frac{dn}{dt}=-\frac{n}{\tau}-kn^2+g.$$

Here $\tau$ is the lifetime and $g$ is the exciton generation rate scaled as $bP_{exc}$, where $P_{exc}$ is the excitation power and $b$ is the generation efficiency. 
Solving the equation for the steady state, the PL intensity is equal to 
$I(P_{exc}) = c_E A \frac{n}{\tau}=c_E A\frac{\sqrt{(\frac{1}{\tau})^2+4kbP_{exc}}-\frac{1}{\tau}}{2k\tau}$.
$c_E$ = 6.8$\times$10$^{-6}$ is the effective collection efficiency, which is obtained from Note 3, and A is the laser spot size (1 $\mu m^2$) for our measurements. Considering the absorption to be 40$\%$, the scaling factor $b$ = 1.45$\times$ 10$^{20} \mu J^{-1}$cm$^{-2}$ in the fitting curve presented in Fig.~S4a and S4b. Fitting parameters $k$ =  5.5$\times$ 10$^{-4}$ cm$^2$/s and $\tau$ = 10 ns.  Both the linear scale (Fig.~S4a) and logarithmic scale (Fig.~S4b) show the quantitative match between the fitting and experimental data. 

Next, we discuss the absorption saturation mechanism \cite{RiveraNComm2015}. For the resonant excitation, the absorption is proportional to $1/(1+P_{exc}/P_{s})$, where $P_s$ is the saturation power. Therefore, the rate equation becomes, 

$$\frac{dn}{dt}=-\frac{n}{\tau}+g. 
$$ $g$ is proportional to $\alpha P_{exc}/(1+P_{exc}/P_{s})$ and $\alpha$ is the linear absorption (11.5$\%$). Making $\frac{dn}{dt}=0$, the steady-state PL intensity is \[I(P_{exc}) = \frac{c_A}{h\nu}\frac{\alpha P_{exc}}{1+P_{exc}/Ps}.\]
Here $c_A$ is the collection efficiency for the absorption saturation model and $h\nu$ is the photon energy (1.72 eV). Using this formula, the fitting results are shown in Fig.~S3c, S3d with $\alpha$ = 11.5$\%$, $c_A$ = 6.8$\times$10$^{-6}$, $Ps$ = 6.3 $\mu$W. Although the linear scale fitting matches well, the logarithmic scale fitting shows that the absorption saturation model deviates from the experimental results in the low power regime, implying that the absorption saturation is not the mechanism for saturation behavior in the experiment.

Comparing these two mechanism, the exciton-exciton annihilation is a more possible explanation for the saturation behavior. Recently, the phase transition from interlayer excitons to charge-separated electron-hole plasmas has been reported in the same system with continuous wave excitation\cite{WangSciAdv2019}, supporting the exciton-exciton annihilation of interlayer excitons.
\begin{figure}[H]
\centering
\includegraphics[scale=0.9]{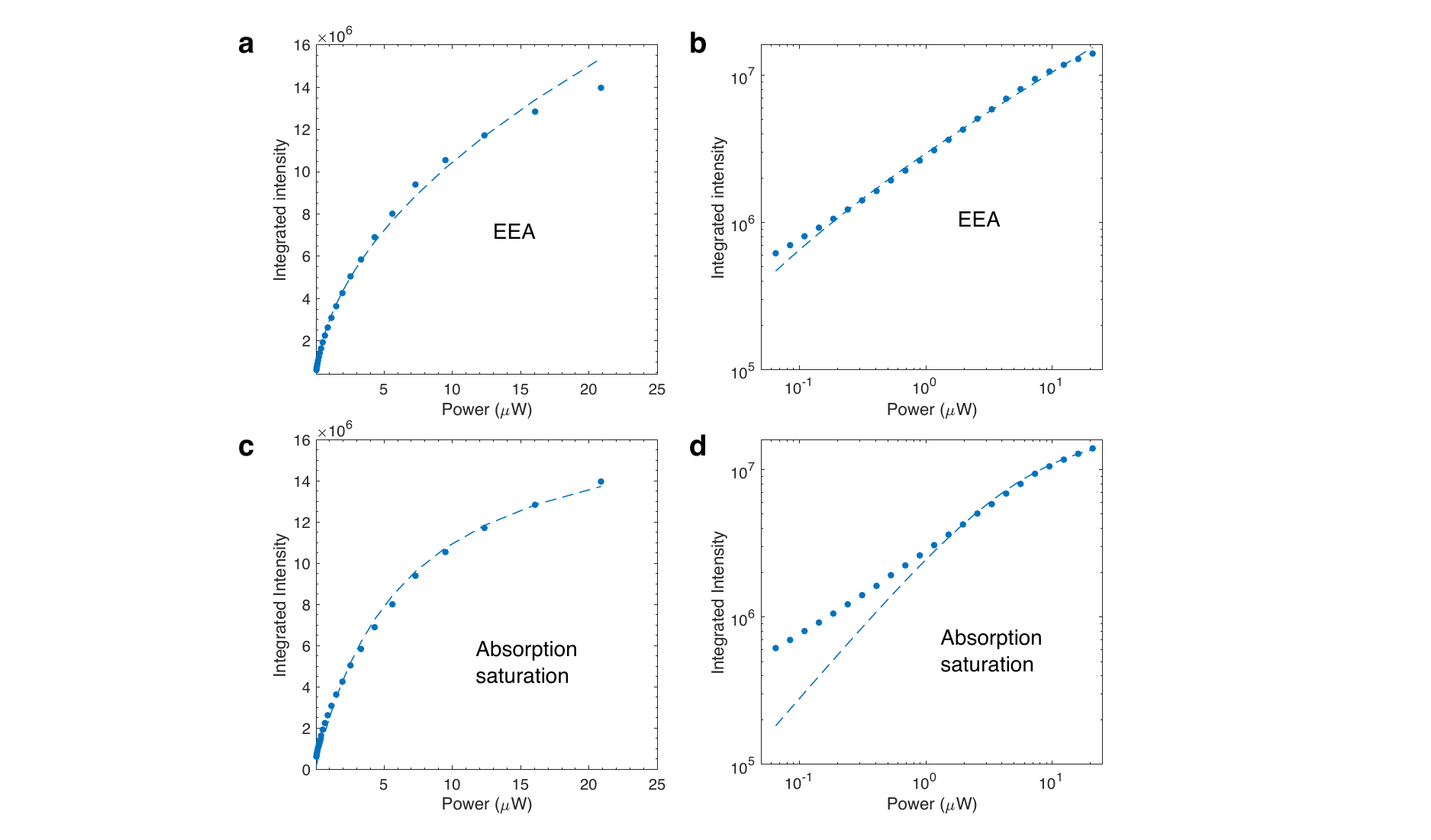}\\
\justify
{\bf Figure S4: Calculated power dependence of the integrated intensity.} {\bf a, b,} Experimental data (blue circles) and fitting (blue dashed lines) using the exciton-exciton annihilation (EEA) model in the linear scale (a) and logarithmic scale (b). {\bf c, d,} Experimental data (blue circles) and fitting (blue dashed lines) using the absorption saturation model in the linear scale (c) and logarithmic scale (d). The logarithmic scale shows that the EEA model fits better than the absorption saturation model, implying the saturation behavior is probably from the EEA process.
\end{figure}


\begin{figure}[H]
\centering
\includegraphics[scale=0.9]{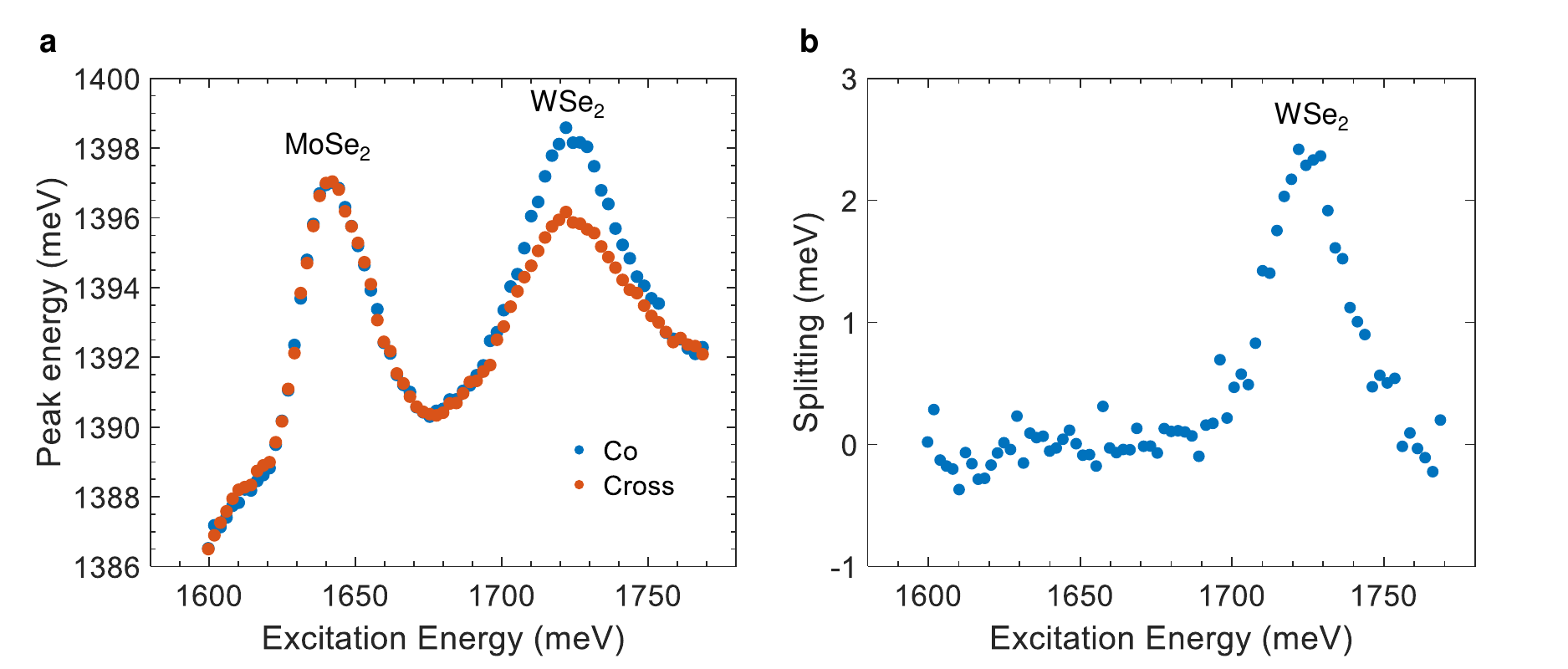}\\
\justify
{\bf Figure S5: Excitation dependence of the peak energy and splitting at circular incidence.} The shifts of interlayer exciton peak in panel {\bf a} show two resonances, corresponding to the MoSe$_2$ ($\sim$1.64 eV) and WSe$_2$ ($\sim$1.72 eV) intralayer exciton energies. The blueshift upon resonant excitation is consistent with the photoluminescence excitation intensity plot in Fig.~1d, providing further evidence for the fact that the exciton energies are closely related to the population of excitons. The peak energy difference or splitting, defined as $E_\mathrm{co}-E_\mathrm{cross}$ is presented in panel {\bf b}, which shows only WSe$_2$ resonance can induce substantial splitting. Combined with Fig.~2a in the main text, it indicates that the imbalance at the WSe$_2$ resonance creates the splitting. Excitation laser is circularly-polarized with incident power $P$ = 2 $\mu$W.
\end{figure}

\begin{figure}[H]
\centering
\includegraphics[scale=0.9]{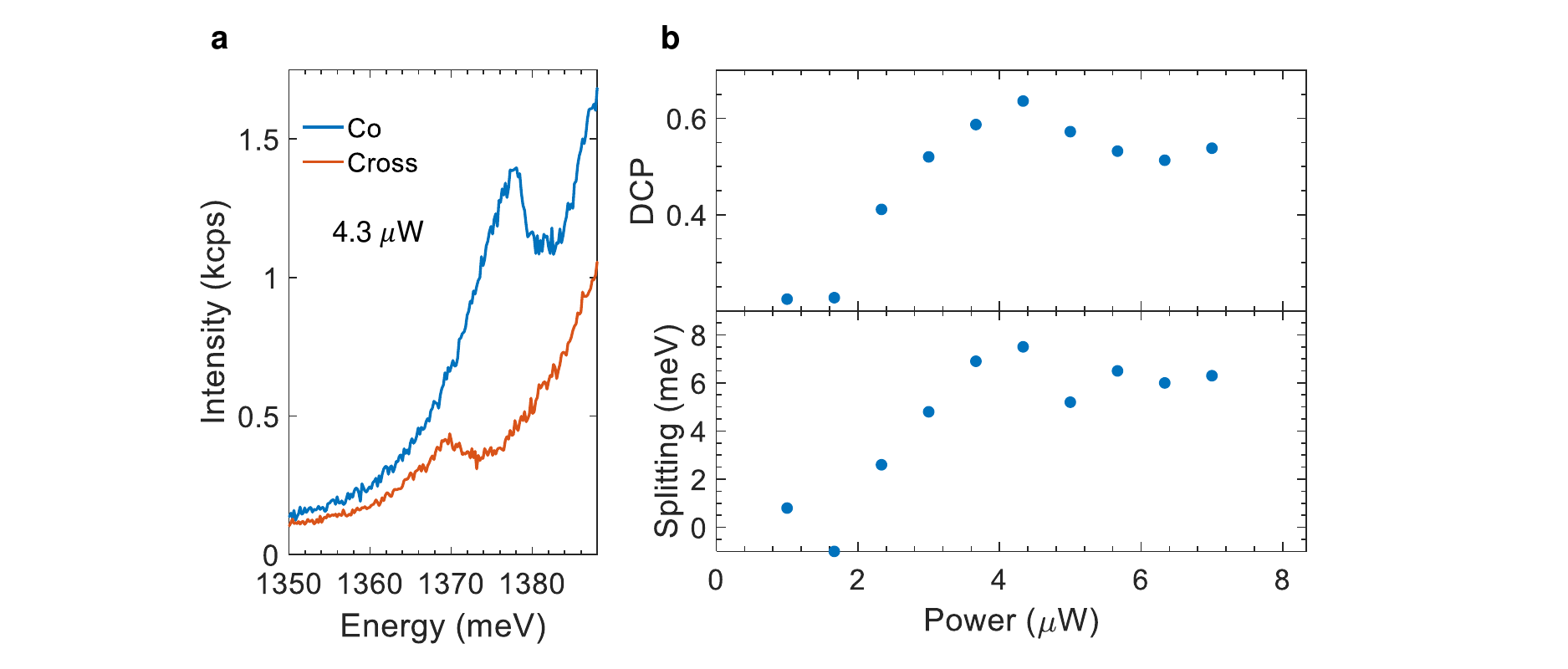}\\
\justify
{\bf Figure S6: Power dependence of the degree of circular polarization and splitting for another interlayer exciton.} {\bf a,} Polarization resolved PL spectra at 4.3 $\mu$W. The co-polarized excitons have higher intensity and higher energy than the cross-polarized excitons, which is the same as the exciton in the main text. {\bf b,} The power-dependent DCP (upper panel) and splitting (bottom panel). Both the DCP and splitting increase with larger power and saturate about 4.3 $\mu$W.  Excitation laser is circularly-polarized, with energy $h\nu$ = 1.72 eV.
\end{figure}

\textbf{Note 3: Calculation of the power-dependent energy splitting.}
\\
Power dependence of the energy splitting can be understood by the density-dependent energy shift from the exciton-exciton interactions\cite{FernandezPRB1996,RiveraScience2016}. The interactions can be separated into two parts, dipole-dipole interaction $U_\mathrm{dd}$ and exchange interaction $U_\mathrm{ex}$. In the heterobilayer system, each exciton carries a permanent dipole moment. Due to the fixed dipole moments from MoSe$_2$ layer to the WSe$_2$ layer, the dipole-dipole interaction is repulsive and induces the blueshift of the interlayer exciton energy. This dipole-dipole interaction is independent of the valley indices of interactive excitons, and thus the blueshift from $U_\mathrm{dd}$ is the same for the two valley excitons $|IX,+\rangle$ and $|IX,-\rangle$, that is, the excitons that couple to $\sigma^+$ and $\sigma^-$ light, respectively. In contrast, the exchange interaction $U_\mathrm{ex}$ only exists between interlayer excitons in the same valley and is a valley-dependent effect. For the valley excitons $|IX,+\rangle$ and $|IX,-\rangle$ with the density $n_{+}$ and $n_{-}$, the mean field description of the exciton-exciton interactions is given by,

\begin{eqnarray}
\Delta E_{+} &=& n_{+}\bar{U}_\mathrm{ex}+ (n_{+}+n_{-})\bar{U}_\mathrm{dd}, \\
\Delta E_{-} &=& n_{-}\bar{U}_\mathrm{ex}+ (n_{+}+n_{-})\bar{U}_\mathrm{dd}.
\end{eqnarray}

where $\Delta E_+$ ($\Delta E_-$) is the energy shift of the $|IX,+\rangle$ ( $|IX,-\rangle$) from exciton-exciton interactions. $\bar{U}_\mathrm{ex}$ and $\bar{U}_\mathrm{dd}$ are the exchange interaction and dipole-dipole interaction in the mean field description. The estimation of  $\bar{U}_\mathrm{ex}$ and $\bar{U}_\mathrm{dd}$ gives $\bar{U}_\mathrm{ex}\sim a_B^2E_b$, and  $\bar{U}_\mathrm{dd}\sim d a_\mathrm{B} E_b$\cite{RiveraScience2016,FernandezPRB1996}, where $a_B$ is the Bohr radius of interlayer exciton, $E_b$ is the binding energy, and the scaling factor  $k \sim$ 1 in the 2D limit. We can obtain the splitting energy by calculating the difference of shift between $|IX,+\rangle$ and $|IX,-\rangle$,

\begin{equation}
\delta E=\Delta E_{+}-\Delta E_{-}=( n_{+}- n_{-})\bar{U}_{ex}.
\end{equation}

This shows that the splitting is proportional to the imbalance of exciton density, $n_{+}- n_{ -}$. At $\sigma^+$-polarized excitation with power P = 5.62 $\mu$W for Fig.~S7, $\Delta E_+$ = 10.0 meV, $\Delta E_-$ = 5.8 meV. Taking $a_B$ = 2 nm, $d$ = 0.7 nm, $E_b$ = 0.2 eV\cite{RiveraScience2016}, we can estimate $ n_+$ = 8.48$\times$10$^{11}$ cm$^{-2}$, and  $ n_-$ = 3.17$\times$10$^{11}$ cm$^{-2}$. The total exciton density $n$ is then 
given by $n_++n_-$ as 1.16$\times$10$^{12}$ cm$^{-2}$. Considering the nano-second lifetime and 10 $\%$ absorption\cite{LiPRB2014}, exciton density is given by $n = \frac{10\%}{A}\times \frac{P}{h\nu}\times \tau$, where A is the laser spot size of 1 $\mu $m$^2$, excitation energy $h\nu$ = 1.72 eV, and $\tau$ is the lifetime. The total density of 1.16$\times$10$^{12}$ cm$^{-2}$ corresponds to the lifetime of 5.7 ns, which is a reasonable value. The measured intensity difference $I_+-I_-$ is 3.6$\times 10^6$, giving the effective collection efficiency $c=(I_+-I_-)/(n_+-n_-)=6.8\times 10^{-6}$, which is used in the Note 2. We note that this effective collection efficiency includes the collection efficiency of our setup including the mode-mismatch due possible $z$-polarized emission of IXs and any non-radiative channels.

 In the main text, we use the degree of circular polarization (DCP) to characterize the imbalance, which is basically DCP = $(n_+- n_-)/(n_++ n_-)$, as $n$ is proportional to intensity $I$. The DCP is proportional to the imbalance, so as to the splitting. In order to connect DCP to the splitting, we add the two energy shift equations (1),(2) and plug into the splitting equation (3) with the DCP, which gives-
 \begin{equation}
 \delta E=(\Delta E_++\Delta E_-)\times \mathrm{DCP}\times\frac{\bar{U}_\mathrm{ex}}{\bar{U}_\mathrm{ex}+2\bar{U}_\mathrm{dd}}=(\Delta E_++\Delta E_-)\times \mathrm{DCP}\times\frac{a_B}{a_B+2d}.
 \end{equation}

Based on the equation, if the excitation is $\sigma^+$-polarized, DCP is positive, and thus the splitting is positive, i.e., $|IX,+\rangle$ has higher energy, while the splitting is flipped for the $\sigma^-$ excitation. Moreover, the splitting $\delta E$ follows the same trend as the DCP, which is consistent with our interpretation throughout the main text.  The calculated energy splitting from the power dependence at B = 0 T is shown in the Fig.~S7. The quantitative agreement between the calculated and experimental splitting at low power before the saturation power validates our model. The deviation starting from the saturation power may come from a phase-space filling blue shift~\cite{KimSciAdv2017}, which is spin-valley independent and is not included in our model.

We note that the strength of exciton-exciton interactions that we find in our sample is $\sim$0.8 $\mu$eV$\mu$m$^2$ and is about an order of magnitude larger than the value reported in monolayer sample~\cite{TanPRX2020,ScuriPRL2018,BarachatiNNano2018}. This strength of interaction $\sim E_b a_B^2$ scales as $a_B /  \varepsilon_r$ where $\varepsilon_r$ is the effective dielectric constant. While $a_B$ is $\sim$10 nm in GaAs quantum wells, due to a larger $\varepsilon_r$, the ratio $a_B /  \varepsilon_r$ is comparable with TMDs\cite{VinaPRB1996,AmandPRB1997}.

\begin{figure}[H]
\centering
\includegraphics[scale=0.9]{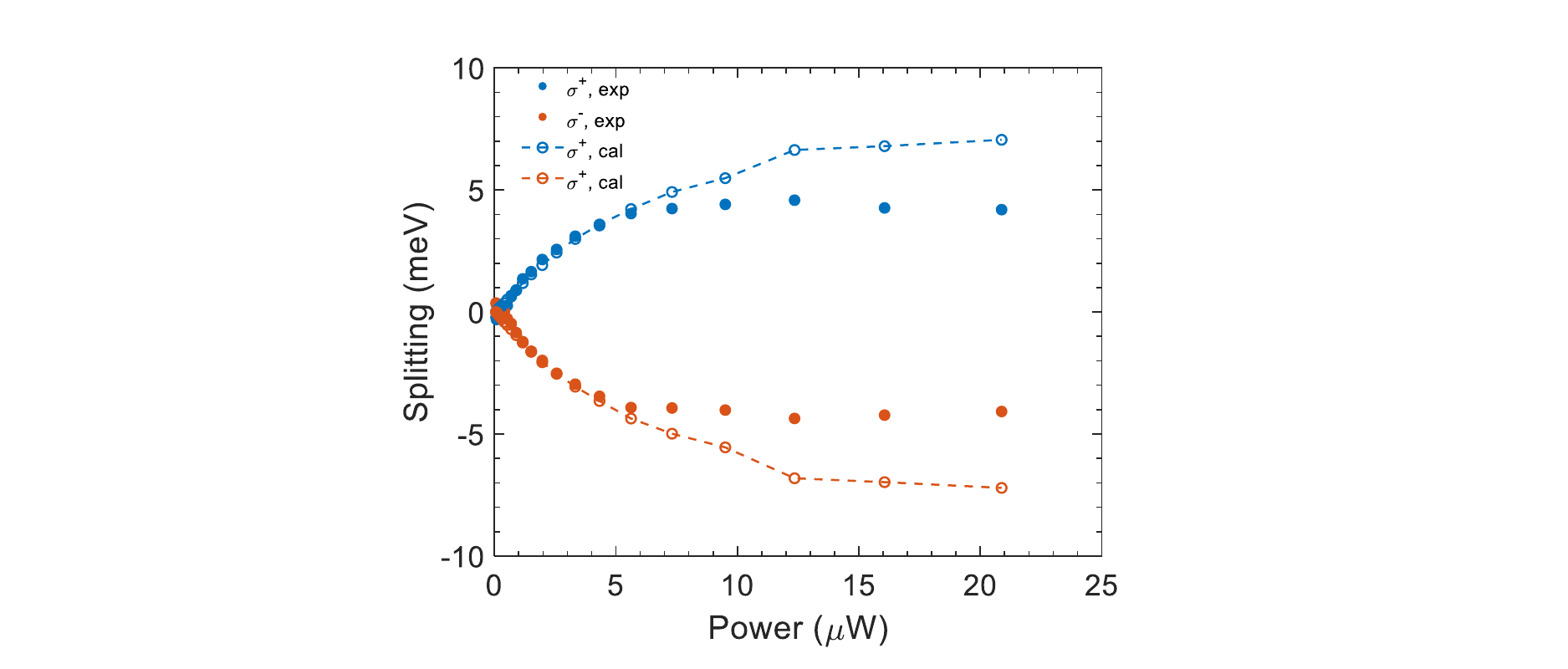}\\
\justify
{\bf Figure S7: Calculated power dependence of the energy splitting.} For both $\sigma^+$ and $\sigma^-$ excitation, the calculated splitting (dash-circle lines) matches the experimental splitting (solid circles) when power is below saturation. Above the saturation power, the experimental splitting is smaller than the calculated splitting, which may be explained by the phase-space filling blue shift independent of spin-valley index. The excitation energy is 1.72 eV for the experimental data.
\end{figure}

\begin{figure}[H]
\centering
\includegraphics[scale=0.9]{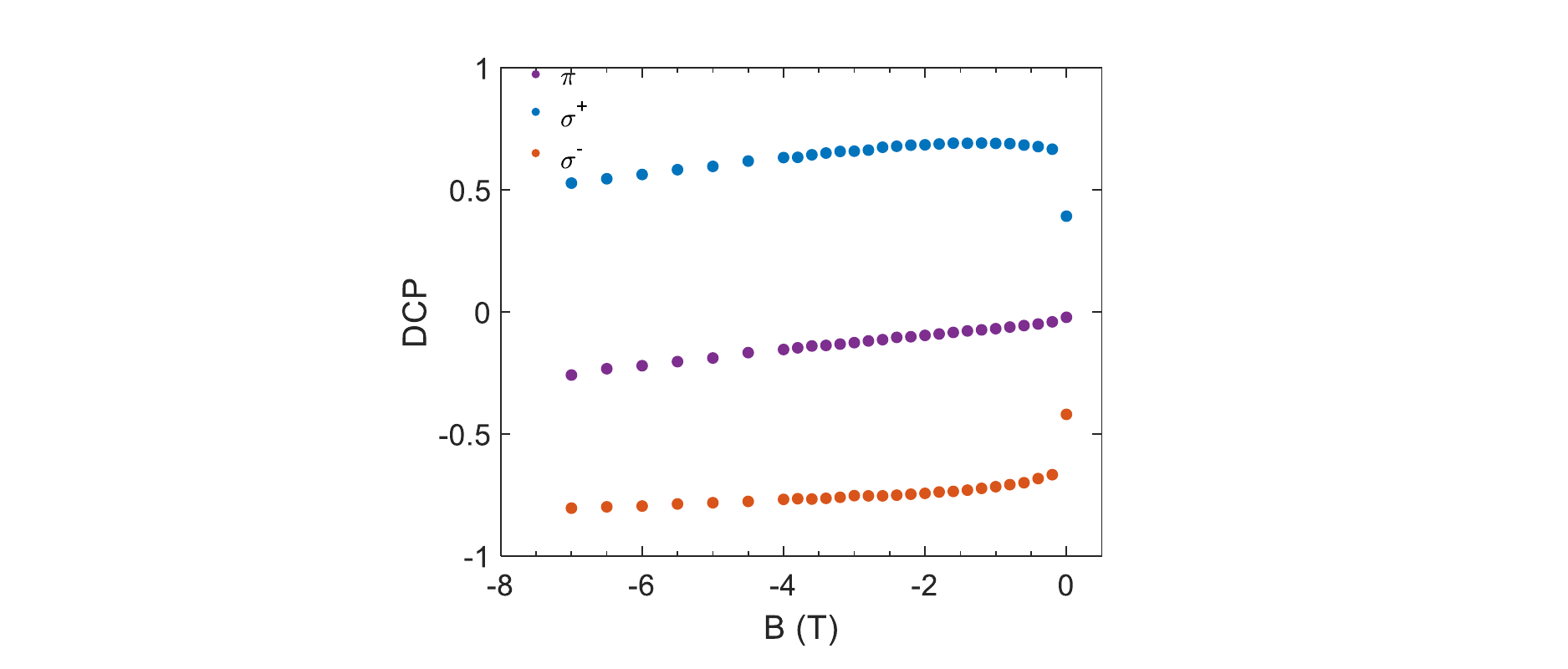}\\
\justify
{\bf Figure S8: Magnetic field dependence of the degree of circular polarization.} Taking the DCP at linear excitation as reference, the DCP of $\sigma^+$ and $\sigma^-$-polarized excitation saturates at B= 0.2 T and then keeps the same, though the magnetic field is increased. Excitation laser energy $h\nu$ is 1.72 eV and incident power is 2 $\mu$W.
\end{figure}

\begin{figure}[H]
\centering
\includegraphics[scale=0.9]{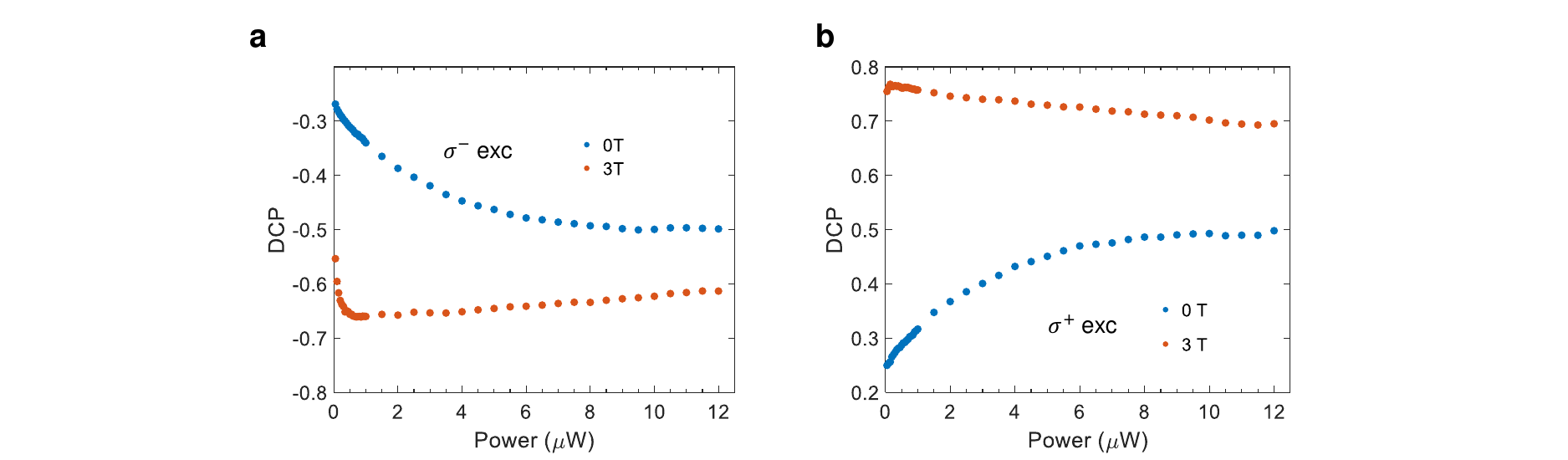}\\
\justify
{\bf Figure S9: Power dependence of the degree of circular polarization at zero and finite magnetic fields (B).}  Power dependence of DCP under B= 0 T and 3 T for $\sigma^-$-polarized excitation (a) and $\sigma^+$-polarized excitation (b). At B = 0 T, the DCP increases  slowly with power and saturates at about 6 $\mu$W, the change of DCP together with the energy shift quantitatively reproduces the splitting as shown in Fig.~S7 (Note 3). However, DCP at B = 3 T saturates at very low power, and decreases slightly when the power increases. Excitation laser is circularly-polarized, with energy $h\nu$ = 1.72 eV.
\end{figure}

\begin{figure}[H]
\centering
\includegraphics[scale=0.9]{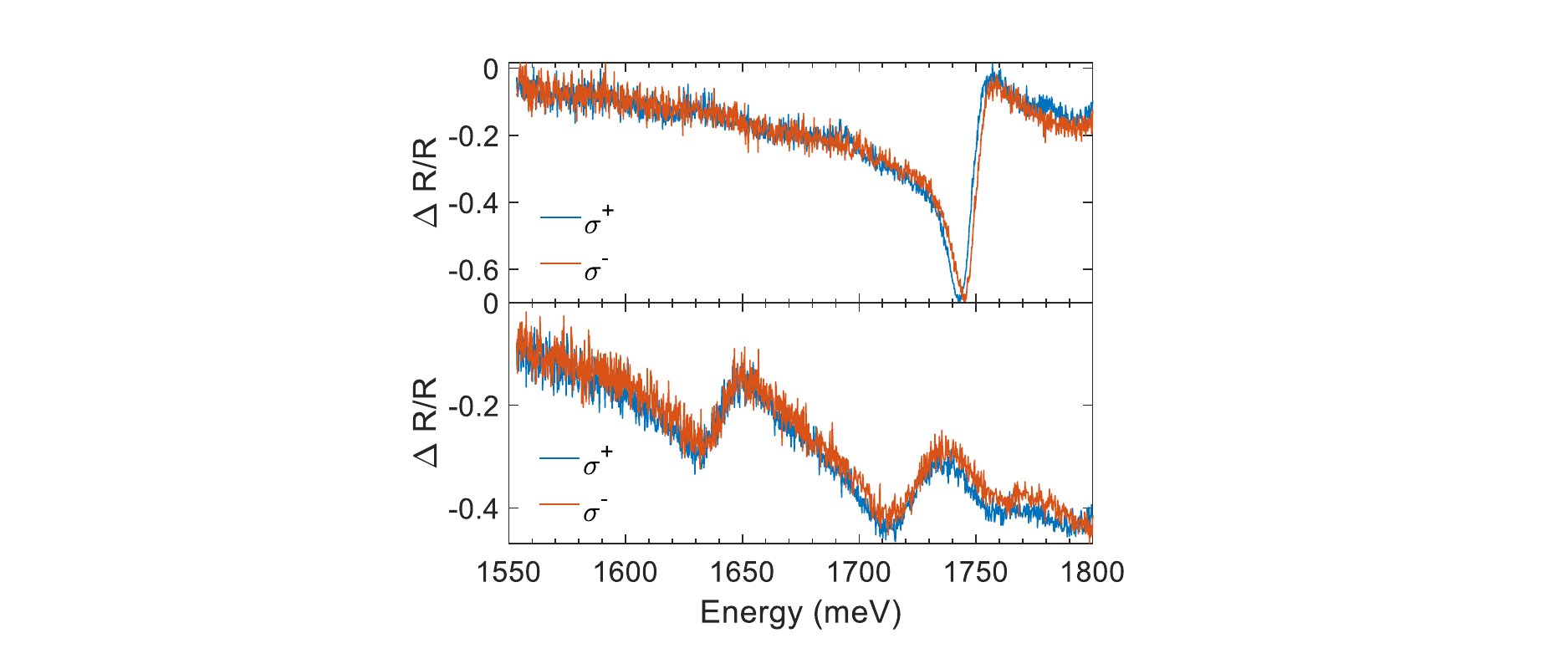}\\
\justify
{\bf Figure S10: Polarization resolved reflectance spectra at linear excitation at B = 7 T.} The upper panel is the monolayer WSe$_2$ reflectance spectra. The monolayer WSe$_2$ absorption energies of $\sigma^+$-polarized excitons (1.7488 eV) and $\sigma^-$-polarized excitons (1.7508 eV) are split by -1.94 meV, which is corresponding to the $\it{g}$ factor of -4.78. The bottom panel is the heterobilayer MoSe$_2$/WSe$_2$ reflectance spectra. The two absorption energies $\sim$ 1.64 eV and 1.72 eV correspond to MoSe$_2$ and WSe$_2$ resonances as shown in the Fig.~1d. The WSe$_2$ resonance in heterobilayer red-shifts from that of the monolayer counterpart, implying the effective interlayer coupling between the MoSe$_2$ and WSe$_2$ layers.
\end{figure}

\textbf{Note 4. Magnetic field dependence of exchange field ($B_{\mathrm{ex}}$).}

Once an external $B$ is applied in the Faraday geometry, the effective magnetic field ($B_{\mathrm{eff}}$) in the hBL sample is the combination of $B$ and the exciton-interaction induced exchange field ($B_{\mathrm{ex}}$), 
\begin{equation}
B_{\mathrm{eff}} = B + B_{\mathrm{ex}}.
\end{equation}
As introduced in the main text, $B_{\mathrm{ex}}$ is nonzero when there is population imbalance between the two species of excitons. $B_{\mathrm{ex}}$ even breaks TRS and induces the valley splitting under zero $B$. The total splitting energy $\delta E_\mathrm{total}$ under external $B$ depends on both $B$ and $B_{\mathrm{ex}}$, 
\begin{equation}
\delta E_\mathrm{total} = \delta E(B) + \delta E(B_{\mathrm{ex}}).
\end{equation}
When $B_{\mathrm{ex}} = 0$ ($\delta E(B_{\mathrm{ex}}) = 0$), $\delta E_\mathrm{total}$ is proportional to $B$ due to the valley Zeeman effect,
\begin{equation}
\delta E_\mathrm{total} = \it{g_0\mu_B}B, 
\end{equation}
with $g_0$ = -13.41 $\pm$ 0.08 (Fig.~3b in the main text). When $B_{\mathrm{ex}} \neq 0$ under larger power circular excitation, $\delta E_\mathrm{total}$ still scales linearly with $B$ (Fig.~4c in the main text) but with a different scaling factor compared to that of the valley Zeeman effect. This result can be empirically captured if $B_{\mathrm{ex}}$ is a function of imbalance $\Delta n$ and $B$, and it is proportional to $B$.
\begin{equation}
B_{\mathrm{ex}}(\Delta n,B)= B_{\mathrm{ex}}(\Delta n,0) \pm  a(P)B, 
\end{equation}
with $B_{\mathrm{ex}}(\Delta n,0)$ responsible for the zero-field splitting and $a(P)$ being constant once the circular excitation power $P$ is fixed. The sign $\pm$ depends on whether $B_{\mathrm{ex}}$ acts in concert or opposition with external $B$. Accordingly, the splitting energy $\delta E_\mathrm{total}$ is given by,
\begin{equation}
\delta E_\mathrm{total} = \mu_B g_0(1\pm a(P)))B+\delta E(B \mathrm{=0~T}).
\end{equation}
The effective $g$ factor $g_\mathrm{eff}$ = $g_0(1 \pm a(P))$. Plugging in the values from Fig.~4c, we obtain $a(7.6 \mu W)$ = 0.25 $\pm$ 0.05. $a(P)$ is an increasing function of $P$ and $\rightarrow 0$ as $P \rightarrow 0$ such that at low power the non-linear behavior discussed in Fig.~4 of the main text is negligible. We note that the linear relation of $B_\mathrm{ex}$ with $B$ in Eq.~8 has been derived on very general grounds and is valid as long as the splitting is linear with $B$. 

The simplest explanation for a linear increase of $B_\mathrm{ex}$ could arise if $\Delta n$ itself increases linearly with $B$. In other words, $B_\mathrm{ex}$ can be expressed in terms of $\Delta n$ as follows -
\begin{equation}
    B_\mathrm{ex}(P,B) \propto \Delta n (P,B) = \Delta n (P,B) \pm b(P) B,
\end{equation}
where $a(P) \propto b(P)$. As shown below, our data supports this hypothesis. From the power-dependence of $a(P)$ or $b(P)$, it appears that the suppression of valley-mixing in the relaxation process from interlayer to IX (as mentioned in the main text) is more effective at larger imbalance. In other words, the surmised valley-mixing, in absence of $B$, is dominant at larger imbalance, making it a density-dependent scattering process.



\textbf{Note 5: Calculation of the magnetic field dependence of the energy splitting.}
\\
We first note that imbalance density, $n_+ - n_-$, is proportional to the difference in intensities of the two circular components $I_+ - I_-$. In order to confirm that the non-linear behavior under circular incident power in the $B$-dependence of Fig.~4 arises from a linear increase in $\Delta n$ with $B$, we must first take into account the linearly polarized data (equivalent to circular power being zero) as reference. Indeed, we performed a similar analysis for the splitting data as well when we subtracted the linear Zeeman splitting in Fig.~4b or plot the parallel guidelines in Fig. 4c. 

To this end, we note that the intensity imbalance for the linearly polarized excitation scales linearly with $B$ with a positive slope, as shown in Fig.~S11a. Hence, the B-dependence of linearly polarized reference is completely antisymmetric in $B$. We use this model-independent fact to decompose the imbalance for the circular incident data into their respective symmetric and antisymmetric components. The circular components are labeled as co- or cross-polarized peaks based on the incident helicity. As shown in Fig.~S11a, the antisymmetric parts of circular components match up with the linear reference. This implies that the only the symmetric component of the co- and cross-polarized data needs to be considered in order to relate $\Delta n$ to splitting.

Fig.~S11b shows the symmetric parts of the co- and cross-polarized peaks which have been offset by 0.02 T data to ignore the effect of dip discussed in the main text. Next, we must convert this data into splitting by using the experimental data of peak energy versus intergrated intensity. In other words,

\begin{equation}
    E_\mathrm{cal} = E_0 + \left( \frac{\Delta E}{\Delta I}\right)\Delta I_\mathrm{sym},
\end{equation}
where $E_\mathrm{cal}$ are the calculated peak energies, $E_0$ is the zero-$B$ energy, $\Delta I_\mathrm{sym}$ is the symmetrized intensity data and $\left( \frac{\Delta E}{\Delta I}\right)$ is obtained from the peak energy versus integrated intensity data. Fig.~S11c shows the mesaured splitting offset by the linearly polarized Zeeman splitting (reference) and the calculated splitting $\Delta E_\mathrm{cal} = E_\mathrm{cal,co} - E_\mathrm{cal,cross}$. We find excellent match between the two, strongly supporting our claim that the nonlinear behavior at high powers arises from the imbalance increasing linearly with $B$. We remark that no external parameters or free parameters have been used in this analysis.

\begin{figure}[H]
\centering
\includegraphics[scale=0.9]{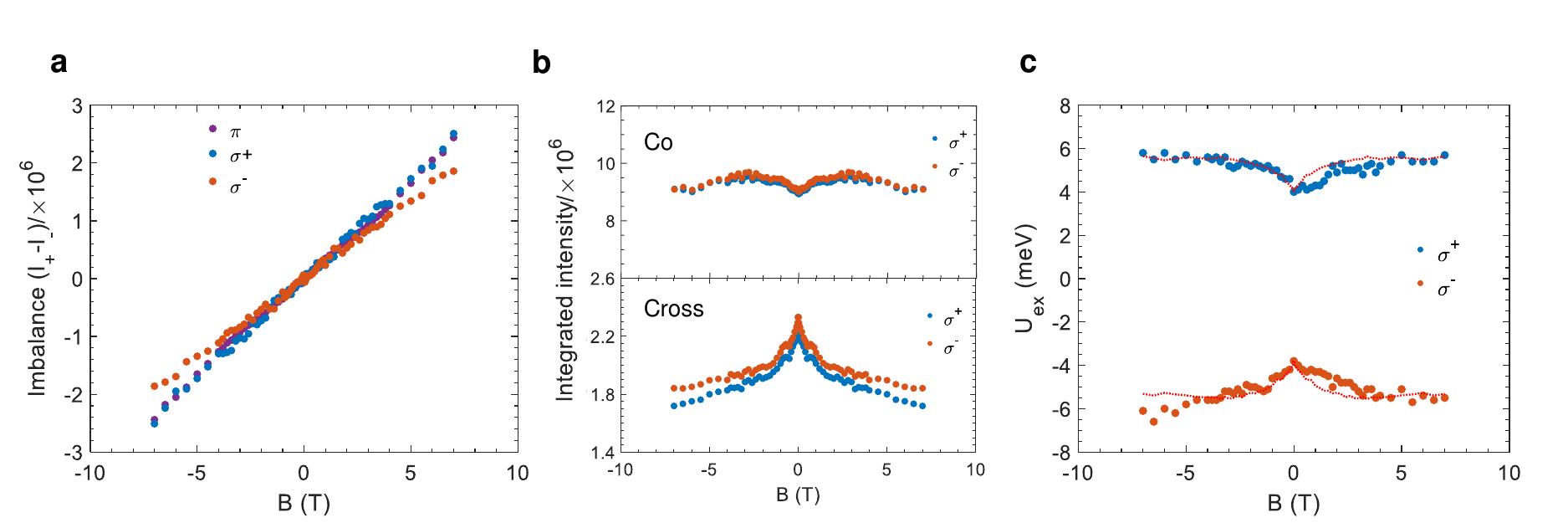}\\
\justify
{\bf Figure S11: Relationship between the imbalance in intensities of the circular components to the splitting} {\bf a,} The linearly-polarized intensity-imbalance data serves as a reference. The antisymmetric parts of the circular data match up almost perfectly with the reference implying only the symmetric part of the circular data contributes to the splitting. {\bf b,} The symmetric parts of the co- and cross-polarized circular data. {\bf c,} A comparison of the calculated splitting (red dashed lines) from the intensity-imbalance data and the measured splitting yields excellent overlap.
\end{figure}


\end{document}